\documentclass[aps,prl,reprint,showpacs,superscriptaddress,longbibliography]{revtex4-2}
\usepackage{mathtools,amssymb,graphicx,units}

\usepackage[plainpages=false,pdfpagelabels,colorlinks=true,linkcolor=red,urlcolor=blue,citecolor=blue,pdftitle={Title},pdfauthor={},pdfdisplaydoctitle=true,pdfduplex=DuplexFlipLongEdge]{hyperref}
\usepackage{verbatim}
\usepackage[english]{babel}
\usepackage{color}
\usepackage{ulem}

\newcommand{\beginsupplement}{%
        \setcounter{table}{0}
        \renewcommand{\thetable}{S\arabic{table}}%
        \setcounter{figure}{0}
        \renewcommand{\thefigure}{S\arabic{figure}}%
}

 \graphicspath{{./fig/}}

\begin{document}
\title{Stark many-body localization on a superconducting quantum processor}
\author{Qiujiang Guo}
\thanks{Those authors contributed equally to this work.}
\affiliation{Interdisciplinary Center for Quantum Information,State Key Laboratory of Modern Optical Instrumentation, and Zhejiang Province Key
    Laboratory of Quantum Technology and Device, Department of Physics, Zhejiang University,
    Hangzhou 310027, China}

\author{Chen Cheng}
\thanks{Those authors contributed equally to this work.}
\affiliation{School of Physical Science and Technology, Lanzhou University, Lanzhou 730000, China}
\affiliation{Beijing Computational Science Research Center, Beijing 100094, China}

\author{Hekang Li}
\thanks{Those authors contributed equally to this work.}
\affiliation{Interdisciplinary Center for Quantum Information,State Key Laboratory of Modern Optical Instrumentation, and Zhejiang Province Key
    Laboratory of Quantum Technology and Device, Department of Physics, Zhejiang University,
    Hangzhou 310027, China}

\author{Shibo Xu}
\affiliation{Interdisciplinary Center for Quantum Information,State Key Laboratory of Modern Optical Instrumentation, and Zhejiang Province Key
    Laboratory of Quantum Technology and Device, Department of Physics, Zhejiang University,
    Hangzhou 310027, China}

\author{Pengfei Zhang}
\affiliation{Interdisciplinary Center for Quantum Information,State Key Laboratory of Modern Optical Instrumentation, and Zhejiang Province Key
    Laboratory of Quantum Technology and Device, Department of Physics, Zhejiang University,
    Hangzhou 310027, China}

\author{Zhen Wang}
\affiliation{Interdisciplinary Center for Quantum Information,State Key Laboratory of Modern Optical Instrumentation, and Zhejiang Province Key
    Laboratory of Quantum Technology and Device, Department of Physics, Zhejiang University,
    Hangzhou 310027, China}

\author{Chao Song}
\affiliation{Interdisciplinary Center for Quantum Information,State Key Laboratory of Modern Optical Instrumentation, and Zhejiang Province Key
    Laboratory of Quantum Technology and Device, Department of Physics, Zhejiang University,
    Hangzhou 310027, China}

\author{Wuxin Liu}
\affiliation{Interdisciplinary Center for Quantum Information,State Key Laboratory of Modern Optical Instrumentation, and Zhejiang Province Key
    Laboratory of Quantum Technology and Device, Department of Physics, Zhejiang University,
    Hangzhou 310027, China}

\author{Wenhui Ren}
\affiliation{Interdisciplinary Center for Quantum Information,State Key Laboratory of Modern Optical Instrumentation, and Zhejiang Province Key
    Laboratory of Quantum Technology and Device, Department of Physics, Zhejiang University,
    Hangzhou 310027, China}

\author{Hang Dong}
\affiliation{Interdisciplinary Center for Quantum Information,State Key Laboratory of Modern Optical Instrumentation, and Zhejiang Province Key
    Laboratory of Quantum Technology and Device, Department of Physics, Zhejiang University,
    Hangzhou 310027, China}

\author{Rubem Mondaini}
\email{rmondaini@csrc.ac.cn}
\affiliation{Beijing Computational Science Research Center, Beijing 100094, China}

\author{H. Wang}
\email{hhwang@zju.edu.cn}
\affiliation{Interdisciplinary Center for Quantum Information,State Key Laboratory of Modern Optical Instrumentation, and Zhejiang Province Key
    Laboratory of Quantum Technology and Device, Department of Physics, Zhejiang University,
    Hangzhou 310027, China}

\begin{abstract}
  Quantum emulators, owing to their large degree of tunability and control, allow the observation of fine aspects of closed quantum many-body systems, as either the regime where thermalization takes place or when it is halted by the presence of disorder. The latter, dubbed many-body localization (MBL) phenomenon, describes the non-ergodic behavior that is dynamically identified by the preservation of local information and slow entanglement growth. Here, we provide a precise observation of this same phenomenology in the case the onsite energy landscape is not disordered, but rather linearly varied, emulating the Stark MBL. To this end, we construct a quantum device composed of thirty-two superconducting qubits,  faithfully reproducing the relaxation dynamics of a non-integrable spin model. Our results describe the real-time evolution at sizes that surpass what is currently attainable by exact simulations in classical computers, signaling the onset of quantum advantage, thus bridging the way for quantum computation as a resource for solving out-of-equilibrium many-body problems.
\end{abstract}

\maketitle

A fundamental characteristic of much sought-after quantum computers, setting them apart from digital classical computers, is their ability of simulating the dynamics of highly entangled many-particle quantum systems~\cite{Preskill2018}. As a consequence, emulating the non-equilibrium  dynamics of many-body systems represents one of the greatest strengths of quantum simulators~\cite{Altman2019}. In the pursuit of building a fully programmable, and fault-tolerant, digital quantum processor, i.e., a flexible gate-based quantum circuit, other \textit{hybrid} (and simpler) platforms have been thriving. Among those, are the superconducting circuits, which by featuring single- and two-qubit gates, with further in-situ knobs to program various system Hamiltonians, represent a quantum \textit{analog} simulator with the potential to bridge the gap between the most advanced existing supercomputers and the yet elusive digital quantum computer~\cite{Georgescu2014,Neill2018}. Here, we provide a concrete example of such quantum advantage, by emulating the dynamics of a quantum system exhibiting both thermalization and its breakdown, depending on the strength of a nonrandom potential, for system sizes that \textit{exact} methods in classical computers cannot compete with.

In particular, the advantage is clearly highlighted by a quick comparison with similar effort that would be demanded on standard computers. For example, for a 32-qubits device, storing the information related to the excitation preserving wave-function requires memory capacities slightly shy of 10 gigabytes. It certainly does not sound dramatic given the current computational capabilities, but to exactly obtain such wave-function, governed by the unitary evolution of some Hamiltonian of interest, \textit{all} of its eigenstates may possess a finite contribution. Thus, the 10 gigabytes requirement needs to be multiplied by the size of the corresponding Hilbert space, shooting up the needed memory to approximately 3 exabytes, more than 500 times of what is available at the largest supercomputer to date~\cite{Fugaku}. Not to mention the computational time associated to obtaining the eigenstates themselves, with algorithms that scale polynomially with the size of the Hilbert space, ultimately rendering the overall computation unpractical \footnote{We estimate that under a perfect parallelization scenario, that is, with no overheads and bandwidth bottlenecks, the distributed diagonalization of such Hamiltonian would require approximately 295 years in the Fugaku supercomputer when using 32 qubits and over 7 months for $N=29$.}.

In the specific isolated generic quantum system we are interested in, the eigenstate thermalization hypothesis~\cite{Deutsch1991,Srednicki1994} provides a framework, based on the random matrix theory and quantum chaos~\cite{Haake_book}, that explains how the expectation value of physical observables at long times gets reconciled with their corresponding thermodynamic averages~\cite{Rigol2008,Alessio2016}. As a result, the system's  out-of-equilibrium evolution progresses in a manner to scramble local information encoded in the initialization over time, erasing memory of the initial conditions when approaching equilibration, even if unaided by an external reservoir~\cite{Clos2016,Neill2016,Kaufman2016}.

Nonetheless, this generic scenario can break down in the presence of added ingredients, ensuing non-ergodic behavior. The most common one is quenched disorder, wherein by the introduction of randomness on the Hamiltonian of interest, thermalization can be prevented. This is manifested by the localization of the wave functions in Hilbert space, which, in physical terms, results in the halting of mass and energy transport~\cite{Nandkishore2015, Altman2018, Abanin2019}. Termed many-body localization (MBL), it has its roots on the non-ergodicity of noninteracting particles in disordered environments, initially introduced by P. W. Anderson~\cite{Anderson1958}. Analytical~\cite{Basko2006, Imbrie2016}, numerical~\cite{Znidaric2008, Pal2010, Jonas2014, Luitz2015, Mondaini2015} and experimental~\cite{Schreiber2015, Choi2016, Smith2016, Luschen2017, Roushan2017, Xu2018, Wei2018, Kohlert2019, Rispoli2019, Guo2020} evidence  indicates though its persistence in the presence of interactions, if disorder is sufficiently large.

Further theoretical support also pointed out different mechanisms in which \textit{clean} systems, i.e., without an explicit random component, can either display pre-thermalization or localization akin to the MBL phenomenon~\cite{Grover14, Yao16, Smith17}. Among them, one is related to the many-body version of a well known single-particle localization process, the Wannier-Stark localization~\cite{Wannier1960, Wannier1962}, wherein particles in a lattice, subjected to an extra linear potential, become localized, displaying quasi-exponentially localized wave functions~\cite{Schulz2019}. If including interactions, recent numerical studies~\cite{van_Nieuwenburg2019, Schulz2019, Taylor2020, Yao2020} have suggested the absence of thermalization, with indicators precisely similar to the standard MBL, without relying, however, on the original $\ell$-bits picture that explains the appearance of local integrals of motion emerging at strong disorder values.

In the language of quantum chaos, our results demonstrate the transition from non-integrability to an emergent one, as the strength of the Stark potential is enhanced, intrinsically related to a fragmentation of the associated Hilbert space~\cite{Sala20}. Deep in the non-ergodic phase, long-lived Bloch oscillations are re-identified, reemphasizing the non-thermal aspects of our emulation, and its connections with other experimental indications in cold atoms~\cite{Meinert2014}. Furthermore, from a technological viewpoint, the argument that the MBL phenomenon may be potentially used as a building block for a quantum memory device, becomes more compelling if no explicit disorder is involved, but rather a precise (and often reproducible) linear variation of the energies. The experimental observation of such phenomenon, Stark many-body localization, is one of the main results of our study.

\begin{figure*}[htbp]
\includegraphics[width=1.5\columnwidth]{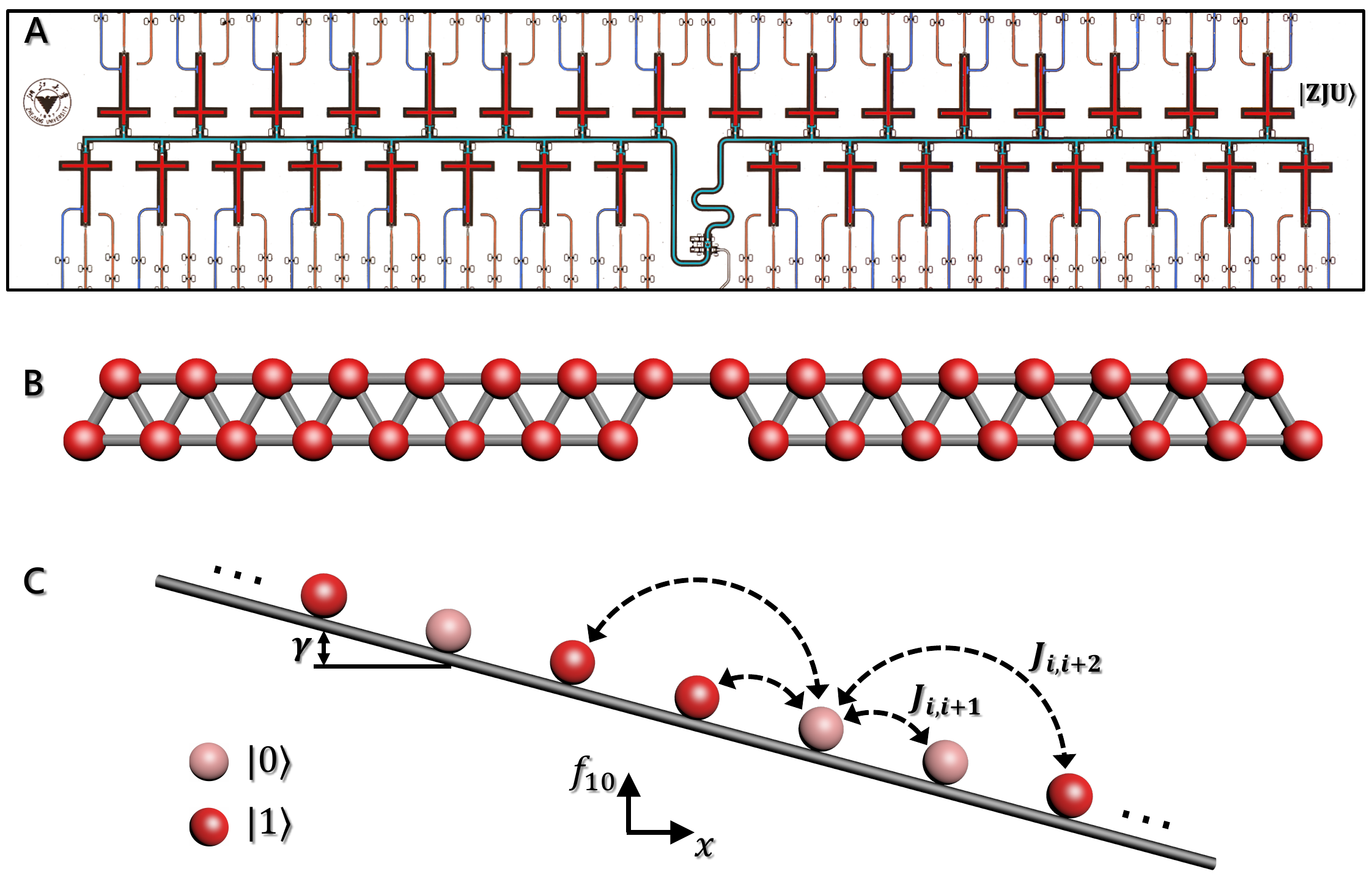}
\caption{\textbf{Quantum processor and schematics of the emulated Hamiltonian.} (\textbf{A}) Optical micrograph of the 32-qubits device, with post-added colors to easily identify the critical circuit elements, as the resonator at the center, control lines and qubits. (\textbf{B}) The schematic representation of the device, with spheres depicting the qubits, and the connecting lines their coupling stemming from the combination of direct- and resonator mediated couplings (See SM for the specific values~\cite{SM}). (\textbf{C}) A cartoon representation of the Hamiltonian, with a linear potential along the device. This is created by the superposition of the homogeneous offset frequency $|\Delta|$ from the resonator and a second small detuning by a controllable value $\gamma$ for consecutive qubits along the $x$-axis, emulating the Stark potential. In the simulations we investigated, half of the qubits are initialized at the excited state $|1\rangle$ via $\pi$-pulses, wherein the coherent dynamics is followed, with subsequent measurements occurring at times $t$ after the initialization. When those occur, all qubits are tuned to their respective readout frequencies for simultaneous multi-qubit state measurement.
}
\label{fig:1}
\end{figure*}

\paragraph{Experimental platform and protocol.---}
We construct an $N=32$ qubit quantum analog processor, using transmon superconducting qubits. With the aid of individual control lines, we program 29 of them, with initializations promoted via microwave photon excitations~\cite{SM}. Their coupling, a mixture of direct and resonator mediated one, describes the effective Hamiltonian~\cite{Xu2018, Song2019}:
\begin{equation}
\label{eq:Ham}
\frac{H}{\hbar} = \sum_{\{i,j\}\in N} J_{ij}\left({\sigma^+_{i} \sigma_{j}^- + \sigma^-_{i} \sigma_{j}^+}\right)
  + \sum_{j\in N} W_j \sigma^+_{j} \sigma_{j}^-.
\end{equation}
Here, $\sigma^+_{i}$ ($\sigma^-_{i}$) is the raising (lowering) operator for qubit $Q_i$, and the first term runs at pairs of qubits $Q_i$ and $Q_j$. Their disposition, accompanied by the qubit-qubit engineered couplings $J_{ij}$, is made such that they emulate a typical non-integrable spin-1/2 model on a triangular ladder (or equivalently, a chain with nearest and next-nearest exchange terms, see Fig.~\ref{fig:1}). Combined with a local adjustment of the resonant frequency for each qubit~\cite{SM}, the second term in \eqref{eq:Ham}, it flexibly allows the construction of numerous potential landscapes $\{W_j\}$, e.g., a linear one, $W_j = -j\gamma$, mimicking a Stark term.

The experiment is performed via the dynamical characterization after an initialization of the qubits, by preparing initial product states via $\pi$-pulse excitation on $N_\bullet$ selected ones, while keeping $N_\circ$-qubits in their ground state. We probe the largest Hilbert space $\cal N$ with a given conserved number of excitations, by choosing $N_\bullet = N/2$, thus amounting to a total of ${\cal N} = 77,558,760$ states. For a finite Stark potential $\gamma$, we select $k=20$ initial states $|\Psi_0\rangle$, carefully chosen to display associated energies $E=\langle\Psi_0| H|\Psi_0\rangle$ residing close to the center of the eigenspectrum of \eqref{eq:Ham} (See SM~\cite{SM}). In the absence of disorder, the average of the dynamical observables for different initializations, and repetitions for each of them, hence constitutes our \textit{statistical} averaging.

\begin{figure*}[htbp]
\includegraphics[width=2\columnwidth]{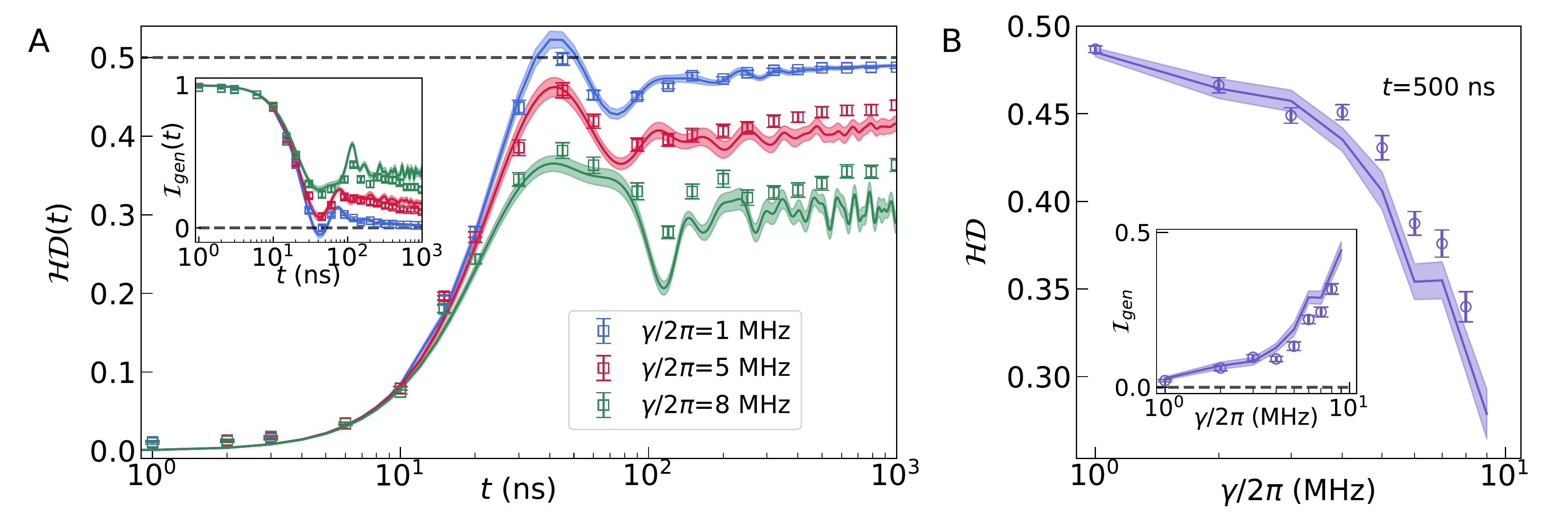}
\caption{\textbf{Relaxation dynamics of few-body observables.} (\textbf{A}) Time-dependent Hamming distance, $\cal HD$, with growing Stark potential $\gamma$, when using 29 qubits of the device depicted in Fig.~\ref{fig:1}(\textbf{A}). Averages are taken among $k=20$ initial states close to the center of the eigenspectrum $\varepsilon \simeq 0.5$. (\textbf{B}) Compilation of Hamming distances at long times; insets in (\textbf{A}) and (\textbf{B}) display similar analysis for the generalized imbalance $\cal I$ at corresponding $\gamma$-values. All data are extracted close to their equilibration values, $t=500$\ ns. Further comparison of the experimental results (markers) with the ones obtained from numerical simulations (lines) is also shown. These employ the time-dependent variational principle (TDVP)~\cite{Haegeman2011,Yang2020}, in which a matrix product state representation of $|\Psi_t\rangle = e^{-\frac{\rm i}{\hbar}Ht}|\Psi_0\rangle$, bounded by the bond-dimension $\chi$, can be obtained (see SM~\cite{SM}). 
The simulation results  become increasingly more expensive when approaching thermalization, due to the growing $\chi$ necessary to faithfully describe $|\Psi_t\rangle$.
Error bars (experiment) and shaded regions (simulation) denote the standard error of the statistical mean.}
\label{fig:2}
\end{figure*}

\paragraph{Dynamically probing Stark MBL.---} We start by characterizing the onset of Stark MBL with growing potentials $\gamma$ by reporting the time-dependent Hamming distance~\cite{Hauke2015,Smith2016},
\begin{equation}
 {\cal HD}(t) = \frac{1}{2}\left(1 - \frac{1}{N}\sum_{i = 1}^{N}\langle \Psi_0 | \sigma_i^z(t)\sigma_i^z(0)|\Psi_0\rangle \right).
\end{equation}
Here, $\sigma_i^z(t)=e^{{\rm i}\frac{H}{\hbar}t}\sigma_i^z(0)e^{-{\rm i}\frac{H}{\hbar}t}$ is the $z$-Pauli matrix for qubit $Q_i$, and ${\cal HD}$ quantifies the expectation value of the normalized (by the system size) number of excitation flips that have occurred at time $t$ in relation to the initial state $|\Psi_0\rangle$. A necessary condition for ergodicity is that at long times $\langle \sigma_i^z (t)\sigma_i^z(0)\rangle \to \langle \sigma_i^z (t)\rangle\langle \sigma_i^z(0)\rangle$, yielding ${\cal HD}(t)\to0.5$; smaller values thus describe memory of the initial preparations. Figure~\ref{fig:2}(A) displays this time dependence for various Stark potentials. At $\gamma/2\pi=1$\ MHz, the Hamming distance tends to 0.5 at the largest experimental time, $t=1000$\ ns, corresponding to approximately 23 tunneling times (see SM~\cite{SM}); it attests that thermalization is achieved in our device within time scales for which evolution is yet coherent. Conversely, at large potentials, memory of the initial conditions persists, with an equilibration of ${\cal HD}<0.5$. This is the first indication of non-ergodic behavior in our system, similar to the MBL phenomenon~\cite{Smith2016}. A compilation of the Hamming distance values at long-times is shown in Fig.~\ref{fig:2}(B).

In addition, further characterization of the preservation of local information is given by the asymptotic values of the imbalance, commonly used in either cold atom~\cite{Schreiber2015,Choi2016,Luschen2017,Kohlert2019} or superconducting qubit experiments~\cite{Xu2018,Guo2020}. Since our initial product states have a generic local  structure, a generalized imbalance~\cite{Guo2020}, defined as ${\cal I}_{\rm gen} = \sum_{i=1}^{N} \lambda_i\sigma^+_i \sigma^-_i$, where $\lambda_i = 1/N_\bullet \  (-1/N_\circ)$ on the  $i$-th qubit initialized to $|1\rangle$ ($|0\rangle$), captures initial local memory. With this construction, $\langle{\cal I}_{\rm gen}\rangle$ is always 1 for the initial state and, similarly to the Hamming distance, it relaxes to a value denoting ergodicity, $\langle{\cal I}_{\rm gen}(t\to\infty)\rangle=0$, while it remains finite at arbitrarily long times when localization takes place. Insets in Fig.~\ref{fig:2} display the imbalance values at the same parameters used for ${\cal HD}$.

\begin{figure*}[htbp]
\includegraphics[width=1.8\columnwidth]{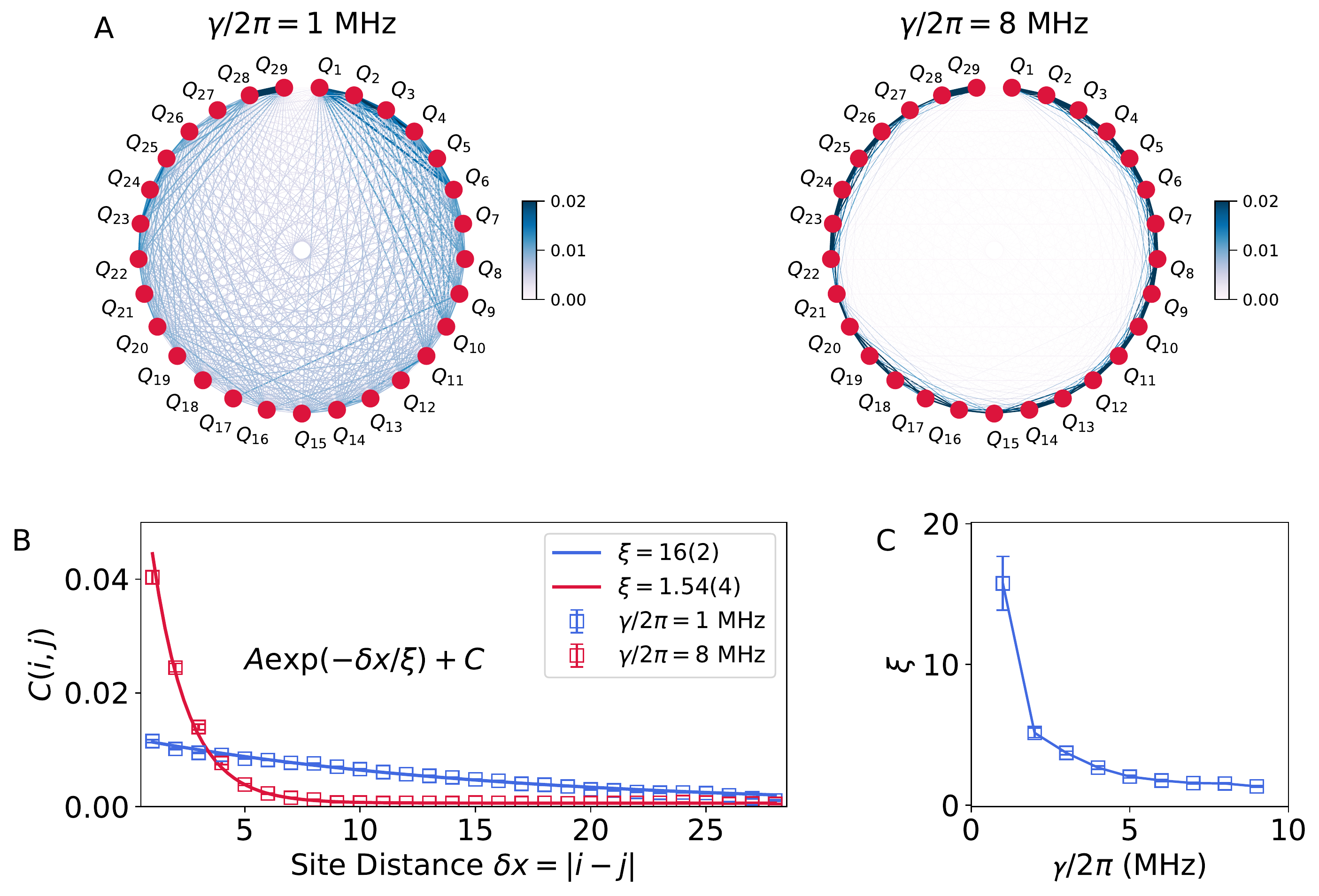}
\caption{\textbf{Ergodicity breaking seen through two-body correlations.} (\textbf{A}) Two-point correlations between all pairs of qubits $(Q_i,Q_j)$ with line color proportional to the correlation strength, at small ($\gamma/2\pi = 1$ MHz) and large ($\gamma/2\pi = 8$ MHz) Stark potential. (\textbf{B}) Averaged two-body correlations $C(i,j)$ across the device as a function of their separation $\delta x \equiv |i-j|$; at small $\gamma$'s, \textit{all} qubits are highly correlated, with a typical correlation length $\xi$ approximately equal to half of the system size. For large tilt potentials, on the other hand, the correlations are much shorter ($\xi\simeq1$), and the system displays a substantially smaller entanglement. (\textbf{C}) The correlation length extracted from the fitting of the functional form $C(i,j) \propto  \exp(-\delta x/\xi) +  C$ as a function of $\gamma$. Error bars in (\textbf{B}) derive from the $k=20$ initial state average, and from the fit procedure in (\textbf{C}); all data is extracted at $t = 500$\ ns, close to the regime where local observables have already equilibrated.}
\label{fig:3}
\end{figure*}

\paragraph{Signifying localization via two-body correlations.---} Supporting indication of the onset of localization is also seen via the scaling behavior of two-point correlation functions~\cite{Neill2018}, defined as
\begin{equation}
 C(i,j) = \left| \langle n_i n_j\rangle - \langle n_i\rangle \langle n_j\rangle \right|,
\end{equation}
where $n_i\equiv \frac{1}{2}(1-\sigma_i^z)$ is the local density operator. They form the building block of quantum information theory measurements, as the quantum mutual information~\cite{De_Tomasi2017}, shown to display an exponential decay with distance~\cite{De_Tomasi2017, Villalonga2020} in the MBL phase (thus bounding the decay of any two-point correlation functions), while slower than exponential in the ergodic phase. Figure~\ref{fig:3}(A) displays measurements for both small ($\gamma/2\pi=1$\ MHz) and large ($\gamma/2\pi = 8$\ MHz) potentials, close to their equilibration times for the local observables ($t = 500$\ ns), after initializations conducted as before (see SM~\cite{SM} for snapshots at different times). A growing Stark potential gives rise to very short-ranged correlations at long-times, mostly restricted to nearest-neighbor qubits. On the other hand, a distinctive signature of the large entanglement is obtained at small values of $\gamma$, with sizeable $C(i,j)$'s spread across the device. By averaging all pairs $\{i,j\}$ of distance-equivalent correlations ($\delta x \equiv |i-j|$), we recover the exponential bound $\propto \exp{(-\delta x/\xi)}$ [Fig.~\ref{fig:3}(B)], where $\xi$ defines the typical correlation length, which quickly drops with larger Stark potentials [Fig.~\ref{fig:3}(C)]. Corrections to this functional form in the localized phase~\cite{Villalonga2020} may yield a slightly different $\xi$, but do not change the highly non-local nature of $C(i,j)$ at small $\gamma$ values.

\paragraph{Stark MBL as opposed to Wannier-Stark localization.---} A last pertinent question is whether the observed localization can be distinguished from Wannier-Stark localization~\cite{Wannier1960}, a typical single-particle phenomenon. In the latter, unitary dynamics also preserves memory of the initial conditions, challenging its differentiation from the argued many-body localization. However, other dynamical metrics can also discern those, as the growth in time of entanglement measures. In the standard MBL phenomenon, based on the $\ell$-bits description of localization, entanglement grows monotonically slow in time as $\propto \log t$~\cite{Bardarson12,Serbyn13}, in contrast to Anderson localization, in which it quickly saturates after initial short dynamics~\cite{Xu2018}.

\begin{figure}[htbp]
\includegraphics[width=1\columnwidth]{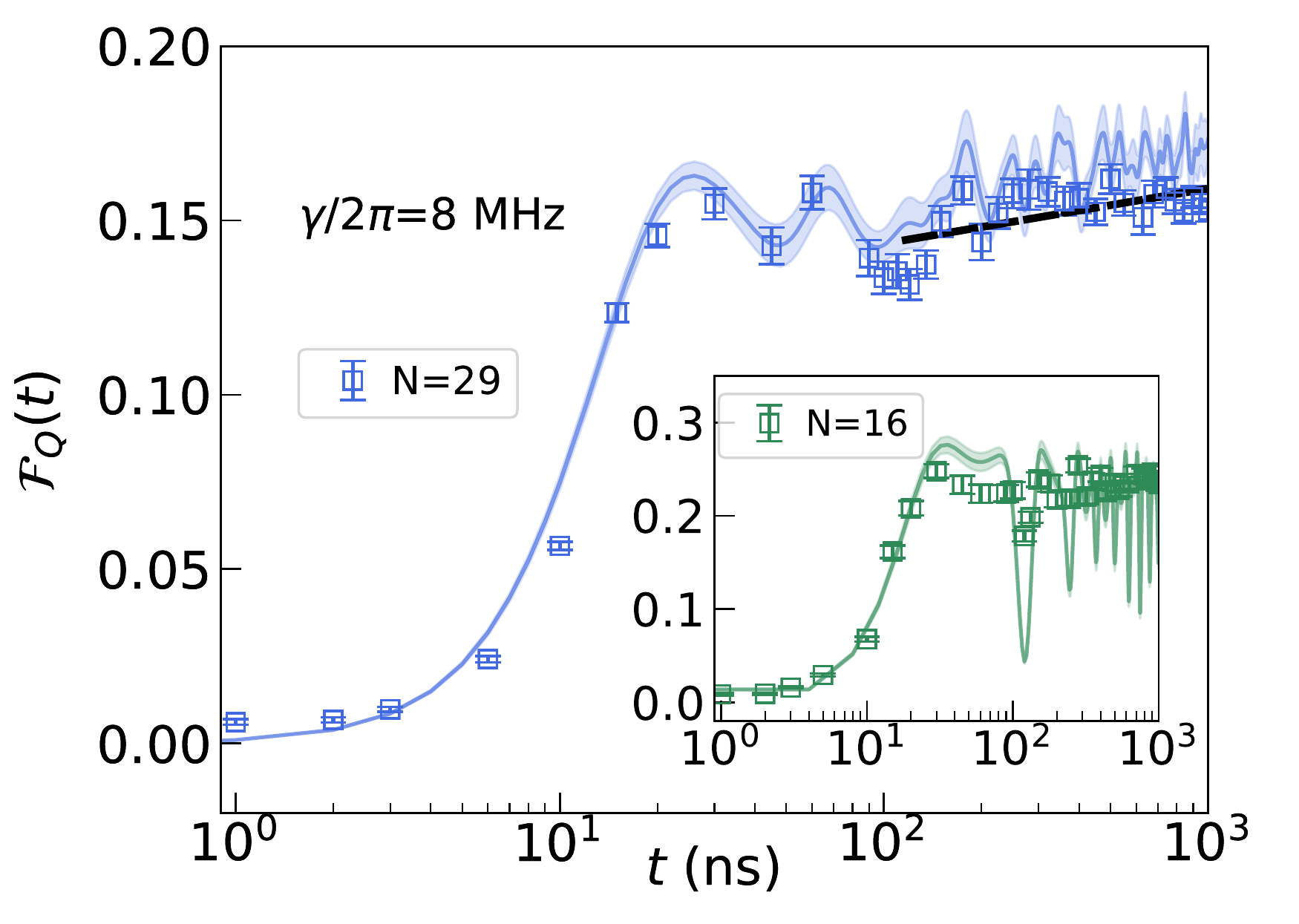}
\caption{\textbf{Entanglement growth -- Stark MBL vs. noninteracting localization.} Time dependence of the experimental quantum Fisher information at large tilt potential ($\gamma/2\pi=8$\ MHz). The dashed line depicts a $\log t$ fit of the experimental data in the range $t\in[100,1000]$\ ns, highlighting the growth at long times in the Stark MBL regime. The inset presents the same when selecting just 16 qubits, such as they feature only nearest-neighbor direct couplings, emulating an integrable Hamiltonian. In all data, averages and error bars stem from standard error of the mean of multiple initializations; markers (lines) denote experimental (numerical simulation) data.}
\label{fig:5}
\end{figure}

\begin{figure}[htbp]
\includegraphics[width=1\columnwidth]{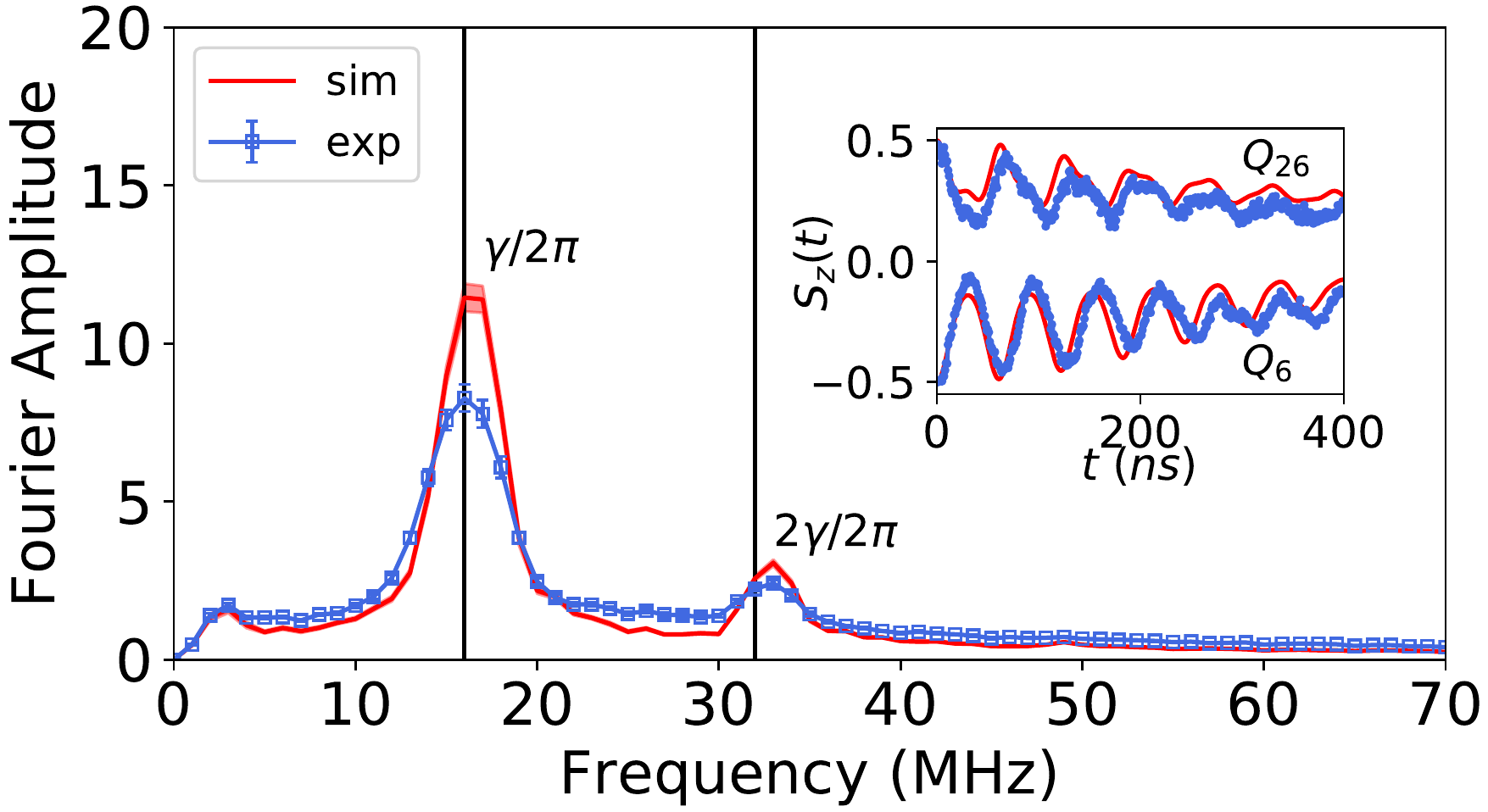}
\caption{\textbf{Bloch oscillations in the Stark MBL phase.} Average Fourier amplitude of the real-time oscillations of $\langle \sigma_i^z\rangle$ for each qubit $Q_i$; inset displays, for example, the dynamics of two qubits, $Q_6$ and $Q_{26}$. These are selected in order to (i) visualize the high amplitude of the oscillations, making the identification of periods more easily discernible, and (ii) amplify the signal to noise ratio; nonetheless, the Fourier amplitudes for the oscillations of \textit{all} qubits are averaged in the main panel. Peaks at $\omega = \gamma/2\pi$ and higher harmonics, describe the Bloch oscillations with frequencies given by the Stark potential, explaining the mechanisms of emerging integrability in the MBL phase. In all data presented, averages and error bars stem from repetitions of a single product state initialization; markers (lines) denote experimental (numerical simulation) data.}
\label{fig:4}
\end{figure}

If we carefully select the qubits in our processor which will be put in a coherent state, one can build an effective linear chain of spins, that cannot be differentiated from non-interacting spinless fermions, after a Jordan-Wigner transformation. Although the half-chain entanglement entropy is an onerous measurement that requires a full quantum state tomography~\cite{Xu2018}, simpler witnesses of entanglement have been used, as the quantum Fisher information (QFI)~\cite{Smith2016, Guo2020}, defined as ${\cal F}_Q=4[\langle {\cal I}_{\rm gen}^2 \rangle - \langle {\cal I}_{\rm gen} \rangle^2]$. 
Figure~\ref{fig:5} contrasts the dynamics of QFI when using 16 of the total number of available qubits, emulating an integrable model, and the original case featuring 29 qubits. The difference in between both results is clear, and goes precisely in confirming the slow logarithm-in-time growth of QFI for the Stark many-body localized case, and the quick saturation for the effective noninteracting Hamiltonian.
\paragraph{Coherent oscillations deep in the Stark MBL regime.---} While the picture of ergodicity breaking is already unequivocal, the tunability available in our quantum platform allows the verification of emergent Bloch oscillations deep in the Stark MBL regime~\cite{Ribeiro2020}. This can be understood from a simple spectral analysis: when $\gamma/\overline{J_{ij}}\gg 1$ ($\overline{J_{ij}}$ is the average of all $J_{ij}$ couplings), the system approaches integrability, and the many-body spectrum is highly degenerate, displaying $\gamma$-spaced energy levels between quasi-degenerate clouds of states~\cite{SM}. Consequently, any few-body observable exhibits oscillations with frequency $\gamma$. Figure~\ref{fig:4} presents a precise test of this prediction by showing the dynamics of a selection of representative quantities, accompanied by a Fourier analysis of the relevant frequencies $\omega$ that describe the real-time oscillations at $\gamma/2\pi = 16$\ MHz. Coherent Bloch fluctuations are observed with $\omega = \gamma$ (and higher harmonics), a result not reproducible at small values of the Stark potential [See Fig.~\ref{fig:2}(A)], where quantum chaotic behavior predominates, and coherent oscillations are only observed at small time scales instead. This confirms the Stark MBL phenomenon we observe, as one related to an emerging integrability at large $\gamma$'s, where the dipole-moment arises as a conserved quantity when $\gamma\to\infty$~\cite{SM}. Yet, an analysis of the ETH indicators establishes that non-ergodic behavior sets in even before fragmentation in the spectrum takes place~\cite{SM}.

\paragraph{Outlook.---} Unlike recent experiments using trapped atoms on an optical square lattice with added tilt potential~\cite{Guardado_Sanchez2020}, the one-dimensional nature of our emulated problem allows us to experimentally observe Stark MBL, as a phenomenon where emergent integrability is manifest within the coherent time scales available, intimately connected to a fragmentation of the Hilbert space. It is important to highlight that arguments stating that thermalization may eventually occur within the dipole-moment conserving subspaces at the regime $\gamma/\overline{J_ij}\gg1$ ~\cite{Khemani20,Moudgalya19}, may render ineffective thermalization provided marginal local variations in the Stark potential, inherent to any realistic experiment, occur. Nonetheless, the MBL phenomenon itself, has been subject to recent scrutiny, and given the number of qubits we have managed to assemble in our quantum simulator, it opens the possibility for testing existing scaling theories of (many-body) localization, complementing results obtained with classical computers at smaller system sizes~\cite{Mace2019}.
\paragraph{Note added.---} At the completing stages of the manuscript, we learned that recent results in one-dimensional cold atom experiments find non-ergodic behavior due to a tilt potential surviving at relevant experimental time-scales~\cite{Scherg2020}, signified by a dynamical memory-preserving scheme, as in our study. Our results verify those, complementing it with entanglement witnesses for a full characterization of the emerging integrability.
\paragraph{\bf Acknowledgments}
\noindent Devices were made at the Micro-Nano Fabrication Center of Zhejiang University.
{\bf Funding:} Supported by National Natural Science
Foundation of China (Grants No. NSAF-U1930402, 11674021, 11974039, 11851110757, 11725419, and 11904145), National Basic
Research Program of China grants (Grants No. 2017YFA0304300 and 2019YFA0308100), and the Zhejiang Province Key Research and Development Program (grant no. 2020C01019). {\bf Author contributions:} C.C. and R.M. proposed the idea; C.C. performed the numerical simulation; Q.G. conducted the experiment; H.L fabricated the device; R.M., C.C., Q.G., and H.W. co-wrote the manuscript; and all authors contributed to the experimental setup, discussions of the results, and development of the manuscript.
{\bf Competing interests:} Authors declare no competing interests.
{\bf Data and materials availability:} All data needed to evaluate the conclusions in the paper are present in the paper or the
 supplementary materials.



\bibliography{stark_sc_qubits}

\clearpage

\onecolumngrid

\begin{center}

{\large \bf Supplementary Materials:
 \\ Stark many-body localization on a superconducting quantum processor }\\

\vspace{0.3cm}

\end{center}

\vspace{0.6cm}

\twocolumngrid

\beginsupplement

This supplementary information describes the experimental device, as its fabrication, engineered qubit couplings, as well as provides supporting information of the ergodicity breakdown via exact numerical calculations on a smaller number of qubits, and other numerical relevant details, as benchmarks of the TDVP method.

\tableofcontents

\section{Device fabrication}
This 32-qubit device was fabricated using the recipe including the following four steps:
\begin{itemize}
\item[1.] Aluminum deposition. 100 nm Al layer is deposited on a 330 $\mu$m one-side polished sapphire substrate using electron beam evaporation.
\item[2.] Etching basic circuitry.  A layer of photoresist (SPR955 7.0) is spin coated on top of the Al layer. Then, patterns for basic circuitry including qubit capacitors,  ground plane, readout resonators, control wires, and bus resonator are defined using a direct-write laser system (DWL66+), which are translated to the Al layer by development and subsequent wet-etching (etchant: 2.38\% TMAH).
\item[3.] Josephson junction deposition and lift-off. Double layers of 500 nm copolymer and 300 nm PMMA  are used to define the  junction area with electron beam lithography (50 kV), followed by development in the MIBK/IPA (1:3) solution. The Al-AlO$_{\rm x}$-Al  Josephson junctions are then deposited using the double-angle electron beam evaporation.
\item[4.] Fabrication of aluminum airbridges. This step is similar to that in Ref.~\cite{Chen2014APL}
\end{itemize}
\section{Device Characterization}
\begin{figure*}[b]
\includegraphics[width=1.8\columnwidth]{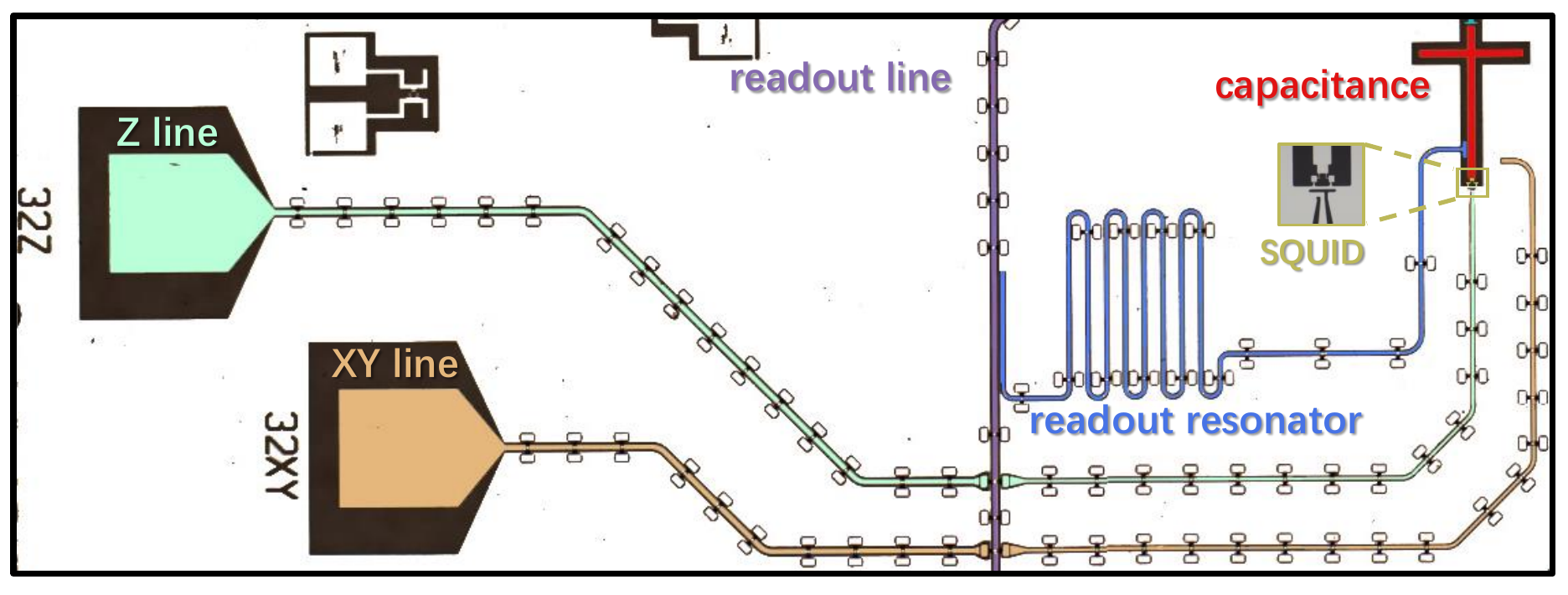}
\caption{Micrograph showing the typical structure of a superconducting qubit. The qubit is a nonlinear oscillator  consisting of a SQUID (yellow region with zoomed-in view in inset), which provides nonlinearity ($\omega_{21}/2\pi-{\omega_{10}}/2\pi\sim -0.33$ GHz) and frequency tunability, and a planar capacitor (red cross). The controllability  is provided with a Z line (cyan), where we supply bias currents to tune qubit frequency and an XY line (brown) for applying microwave pulses to excite $|0\rangle\leftrightarrow|1\rangle$ transition.  Qubit state measurements are implemented by detecting the dispersion of the probe signal through the readout line (purple), which is inductively coupled to a $\lambda/4$ superconducting resonator (blue) whose resonant frequency is qubit state-dependent. In this chip, there are 32 similar qubit modules like this. In order to suppress parasitic modes, different ground planes separated by  coplanar waveguides are connected using air-bridges.}
\label{singleQubitSketch}
\end{figure*}

The device [Fig.~\ref{fig:1}(\textbf{A})] used in this work consists of 32 transmon qubits, and a central bus resonator $\cal{R}$. The typical circuit structure for a qubit, as shown in Fig.~\ref{singleQubitSketch}, is composed of a capacitor, a  superconducting quantum interference device (SQUID) and the additional control circuitry, including a Z line for tuning qubit frequency, an XY line for exciting qubit $|0\rangle\leftrightarrow|1\rangle$ transition, and a  capacitively
coupled $\lambda/4$ coplanar waveguide resonator for dispersive readout. The  typical qubit lifetimes are $\sim$23 $\mu$s for the relaxation time $T_1$ and  $\sim$0.9 $\mu$s for the Ramsey dephasing time, both of which vary depending on the qubit frequency. We note that the true timescales describing the dephasing effects for an \textit{interacting} many-body system, actively coupled, are in practice much longer than the single-qubit Ramsey dephasing times, as has been reported in our previous work using an analogous device~\cite{Guo2020}. The detailed information on qubits' performance is collected in Tab.~\ref{Table S1}. The experimental setup for qubit control and readout can be found in Ref. \cite{Song2019}.

\begin{table*}[!htbp]
    \setlength{\columnsep}{8pt}
	\centering
	{\tabcolsep 0.2in  \begin{tabular}{cccccccccc}
    	\hline
    	\hline
	     &$\omega_{j, {\rm idle}}/2\pi$&$T_{1j, \rm idle}$&$\overline{T}_{1j}$&$\omega_{r,j}/2\pi$&$F_{0,j}$&$F_{1,j}$\\
	        &(GHz)&($\mu$s)&($\mu$s)&($\mu$s)&(MHz)&(GHz)& \\
	\hline
	$Q_1$   & 5.529&$\sim$27&$\sim$29& 6.571& 0.972&0.914\\
	$Q_2$   & 4.260&$\sim$39&$\sim$21& 6.550& 0.929&0.896\\
	$Q_3$   & 5.060&$\sim$19&$\sim$26& 6.600& 0.981&0.915\\
	$Q_4$   & 4.159&$\sim$57&$\sim$29& 6.533& 0.940&0.909\\
	$Q_5$   & 5.681&$\sim$16&$\sim$28& 6.612& 0.970&0.906\\
	$Q_6$   & 4.083&$\sim$42&$\sim$21& 6.504& 0.957&0.916\\
	$Q_7$   & 4.761&$\sim$34&$\sim$29& 6.651& 0.975&0.941\\
	$Q_8$   & 4.294&$\sim$23&$\sim$18& 6.478& 0.962&0.918\\
	$Q_9$   & 4.857&$\sim$27&$\sim$27& 6.466& 0.957&0.921\\
	$Q_{10}$& 5.564&$\sim$17&$\sim$21& 6.721& 0.964&0.851\\
	$Q_{11}$& 5.423&$\sim$20&$\sim$26& 6.444& 0.936&0.904\\
	$Q_{12}$& -    & -      &      - & 6.647& -    &-    \\
	$Q_{13}$& 4.929&$\sim$21&$\sim$23& 6.439& 0.969&0.915\\
	$Q_{14}$& 4.130&$\sim$40&$\sim$24& 6.608& 0.942&0.888\\
	$Q_{15}$& 4.340&$\sim$37&$\sim$28& 6.431& 0.947&0.910\\
	$Q_{16}$& 5.040&$\sim$24&$\sim$20& 6.543& 0.958&0.900\\
	$Q_{17}$& 4.229&$\sim$60&$\sim$22& 6.525& 0.931&0.907\\
	$Q_{18}$& 5.720&$\sim$9 &$\sim$31& 6.469& 0.945&0.869\\
	$Q_{19}$& 5.599&$\sim$25&$\sim$22& 6.531& 0.942&0.919\\
	$Q_{20}$& 4.959&$\sim$16&$\sim$21& 6.529& 0.964&0.932\\
	$Q_{21}$& 3.997&$\sim$34&$\sim$21& 6.512& 0.963&0.931\\
	$Q_{22}$& 4.902&$\sim$29&$\sim$22& 6.594& 0.954&0.892\\
	$Q_{23}$& -    & -      &$\sim$22& 6.528& -    &-    \\
	$Q_{24}$& 5.453&$\sim$29&$\sim$21& 6.650& 0.961&0.900\\
	$Q_{25}$& 4.733&$\sim$28&$\sim$24& 6.708& 0.951&0.920\\
	$Q_{26}$& 4.050&$\sim$28&$\sim$18& 6.393& 0.935&0.912\\
	$Q_{27}$& 4.830&$\sim$31&$\sim$25& 6.661& 0.919&0.889\\
	$Q_{28}$& 5.635&$\sim$15&$\sim$28& 6.439& 0.951&0.872\\
	$Q_{29}$& -    & -      &$\sim$22& 6.597& -    &-    \\
	$Q_{30}$& 4.213&$\sim$30&$\sim$ 6& 6.488& 0.903&0.888\\
	$Q_{31}$& 4.801&$\sim$35&$\sim$19& 6.568& 0.931&0.904\\
	$Q_{32}$& 5.497&$\sim$22&$\sim$24& 6.537& 0.959&0.890\\

		\hline
		\hline
	\end{tabular}}

	\caption{\label{Table S1}\textbf{Typical device performance}. $\omega_{j, {\rm idle}}$ is the $Q_j$'s idle frequency, where the initial states are prepared. $Q_{12}$, $Q_{23}$, $Q_{29}$ are not used, which are placed around their sweet points ($>$ 6.5 GHz) during the whole experiment. The energy relaxation time of $Q_j$ at its idle frequency is denoted as $T_{1j, {\rm idle}}$.  $\overline{T}_{1j, \rm operation}$ describes the averaged energy relaxation time from 5.0 GHz to 5.3 GHz, where qubits interacting with each other for emulating Stark many-body localization physics for different linear potentials. $\omega_{r,j}$ is the resonant frequency of $Q_j$'s readout resonator ($R_j$). The readout fidelities are characterized by the measured probability for state $|0\rangle$ ($|1\rangle$), labeled as $F_{0, j}$~($F_{1, j}$), when $Q_j$ is prepared in state $|0\rangle$ ($|1\rangle$).}

\end{table*}

\section{The physical model for numerics}

In this device, the connectivity $J_{ij}$, in equation \eqref{eq:Ham} of the main text, is provided by both the central bus $\cal{R}$  mediated interactions and the capacitance induced direct couplings. The former are reasonably small due to the large detunings between the bus frequency ($\sim~7$ GHz) and the qubits' interaction frequencies (around 5.15 GHz), so that the capacitance induced direct couplings dominate the system's dynamics. The physical model emulated in the experiment is extracted by systematically measuring the dominant $J_{ij}$ terms, which is used as an input for numerical simulations.

\begin{figure*}[t]
\includegraphics[width=2\columnwidth]{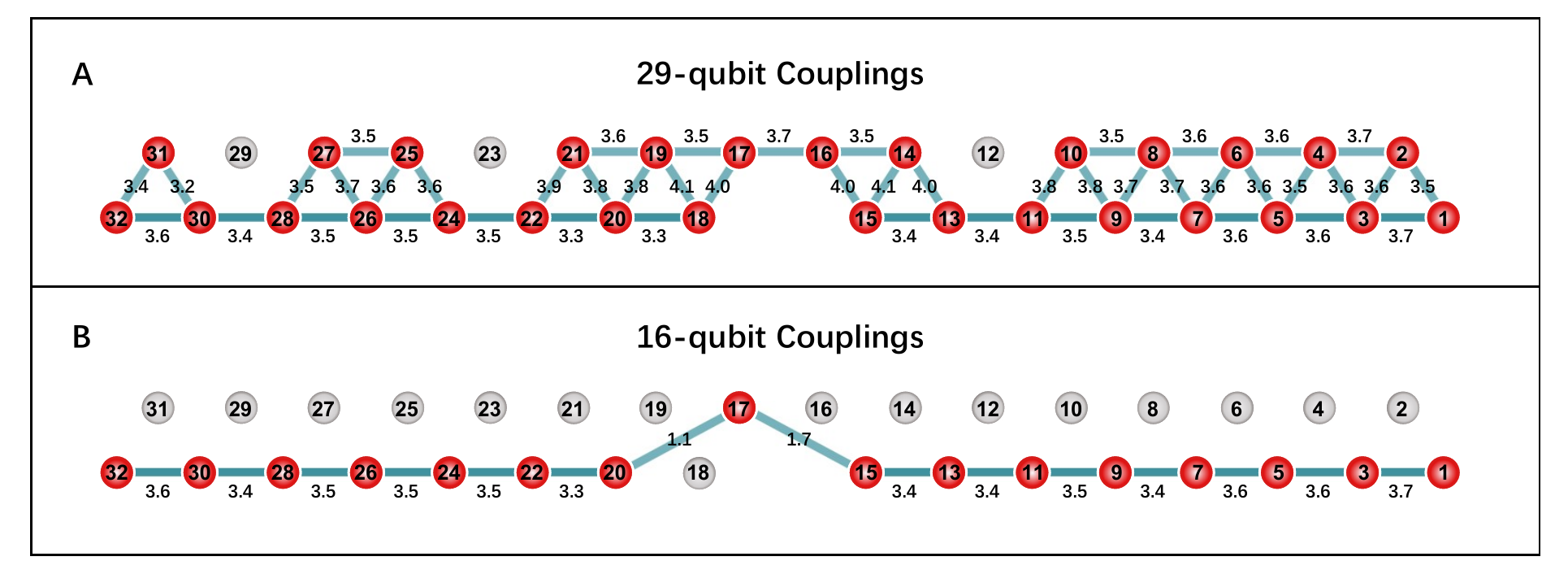}
\caption{\textbf{Coupling topology used for numerical simulations.} (\textbf{A}) The coupling map of the 29 qubits for emulating Stark many-body localization. Red balls with numbers denote the qubits, and cyan lines connecting the balls represent the qubit couplings, with the coupling strengths $J_{ij}/2\pi$ listed. (\textbf{B}) The coupling map describing the effective linear chain of 16 qubits used in characterizing non-interacting localization (See Fig.~\ref{fig:5} of the main text).}
\label{fig:QubitCouplings}
\end{figure*}

At the current stage, three qubits do not function properly, thus, in this work, we use 29 qubits for emulating Stark many-body localization. The coupling topology is shown in Fig.~\ref{fig:QubitCouplings}(\textbf{A}). By using the displayed coupling values, numerical simulations are carried out to benchmark the experiment. Note that qubit indices have been relabeled from $Q_1$ to $Q_{29}$, from right to left, for the selected qubits in  Fig~\ref{fig:QubitCouplings}(\textbf{A}). Figure~\ref{fig:QubitCouplings}(\textbf{B}) shows the coupling topology of 16 qubits, presenting an integrable model, used to quantify Wannier-Stark localization in the main text.

\section{Measurements of Qubit Couplings}
\begin{figure}[h]
\includegraphics[width=8cm]{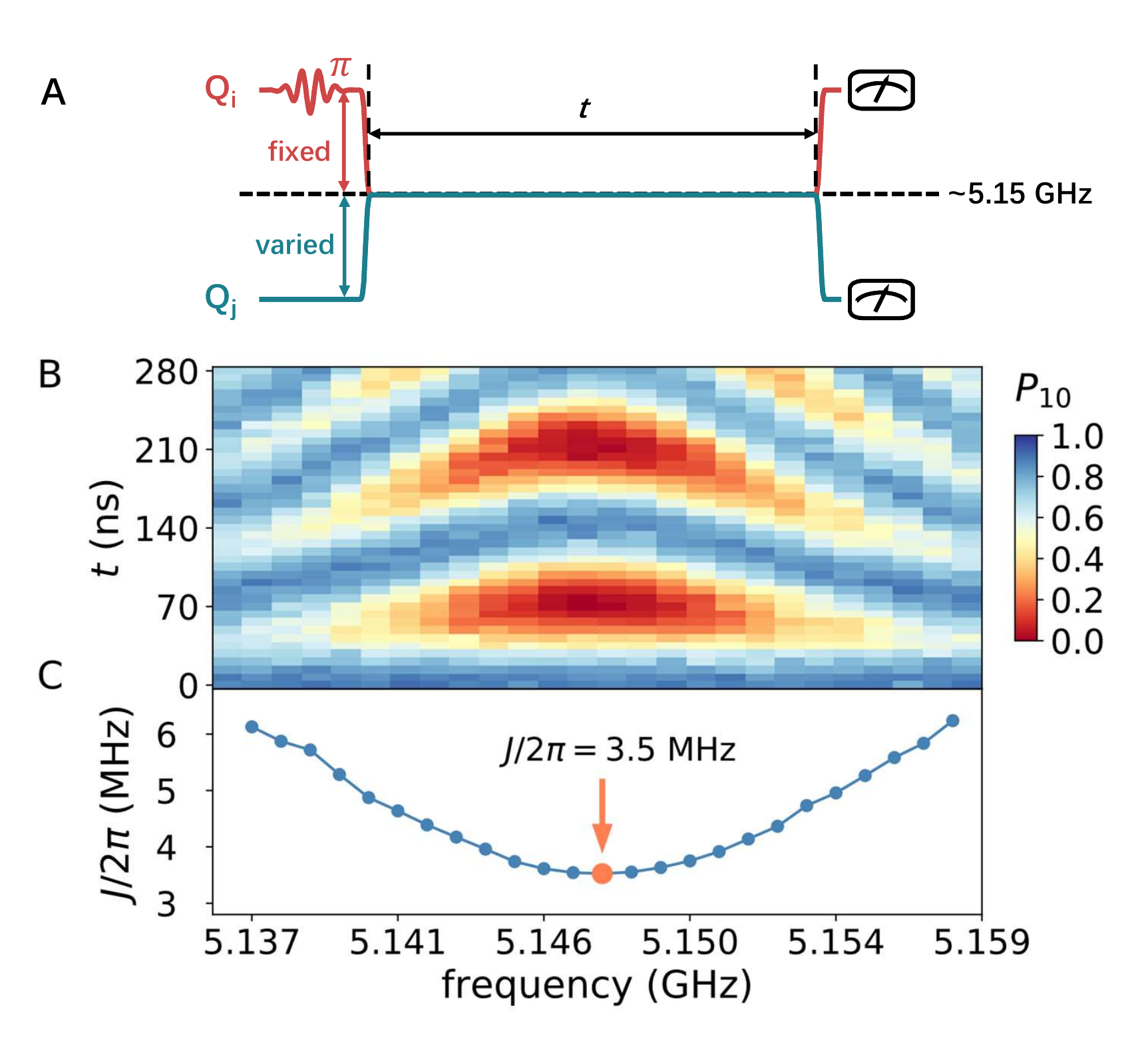}
\caption{\textbf{Measurement of the qubit couplings.} (\textbf{A}) Pulse sequence for calibrating the coupling strength $J_{ij}$ between $Q_i$ and $Q_j$. (\textbf{B}) Experimentally measured swap dynamics for calibrating the coupling strength between $Q_{10}$ and $Q_{8}$. (\textbf{C}) Fitting results using the experimental data in panel (\textbf{B}), where the orange dot indicates the resonant interaction point, with a coupling strength of 3.5 MHz.}
\label{fig:qqSwap}
\end{figure}

 To extract the effective physical model emulated in this experiment, $J_{ij}$ is carefully calibrated by measuring $Q_i$-$Q_j$'s on-resonance energy swap dynamics at the frequency around 5.15 GHz, the center of the operation frequencies of the 29 qubits. The related pulse sequence is shown in Fig. \ref{fig:qqSwap}(\textbf{A}), in which $Q_i$ and  $Q_j$ are prepared to $|10\rangle$ and then, two square pulses are applied simultaneously to tune their frequencies to $\sim$5.15 GHz, with $Q_j$'s frequency being varied. After an interaction time $t$, both qubits are jointly measured to record $P_{10}$ versus $t$. For example, Fig. \ref{fig:qqSwap}(\textbf{B}) shows the experimental data for calibrating the coupling strength between $Q_{10}$ and $Q_{8}$. When the two qubits are on resonance, the oscillation of $P_{10}$ versus $t$ reaches a minimum in frequency, which is twice of the coupling strength $J_{ij}$/2$\pi$. The corresponding fitting results are shown in Fig. \ref{fig:qqSwap}(\textbf{C}), where the minimum point (orange) indicates a coupling strength of 3.5 MHz for $Q_{10}$ and $Q_{8}$.

\section{Synchronization of Z pulses via bus resonator}

For the device with only nearest-neighbor connectivity, the timing offsets between the control lines of different qubits are calibrated by selecting a reference qubit and measuring the relative offsets of the neighboring qubits. The process is done consecutively until all qubits are covered. The issue is that the calibration errors tend to propagate and accumulate as the number of qubits increases. In our device, the central bus resonator could be tuned in frequency to interact with the qubits, providing an optimal medium to calibrate such timing offsets between any pair of qubits directly, which could minimize the accumulation of errors.

\begin{figure}[h]
\includegraphics[width=8.5cm]{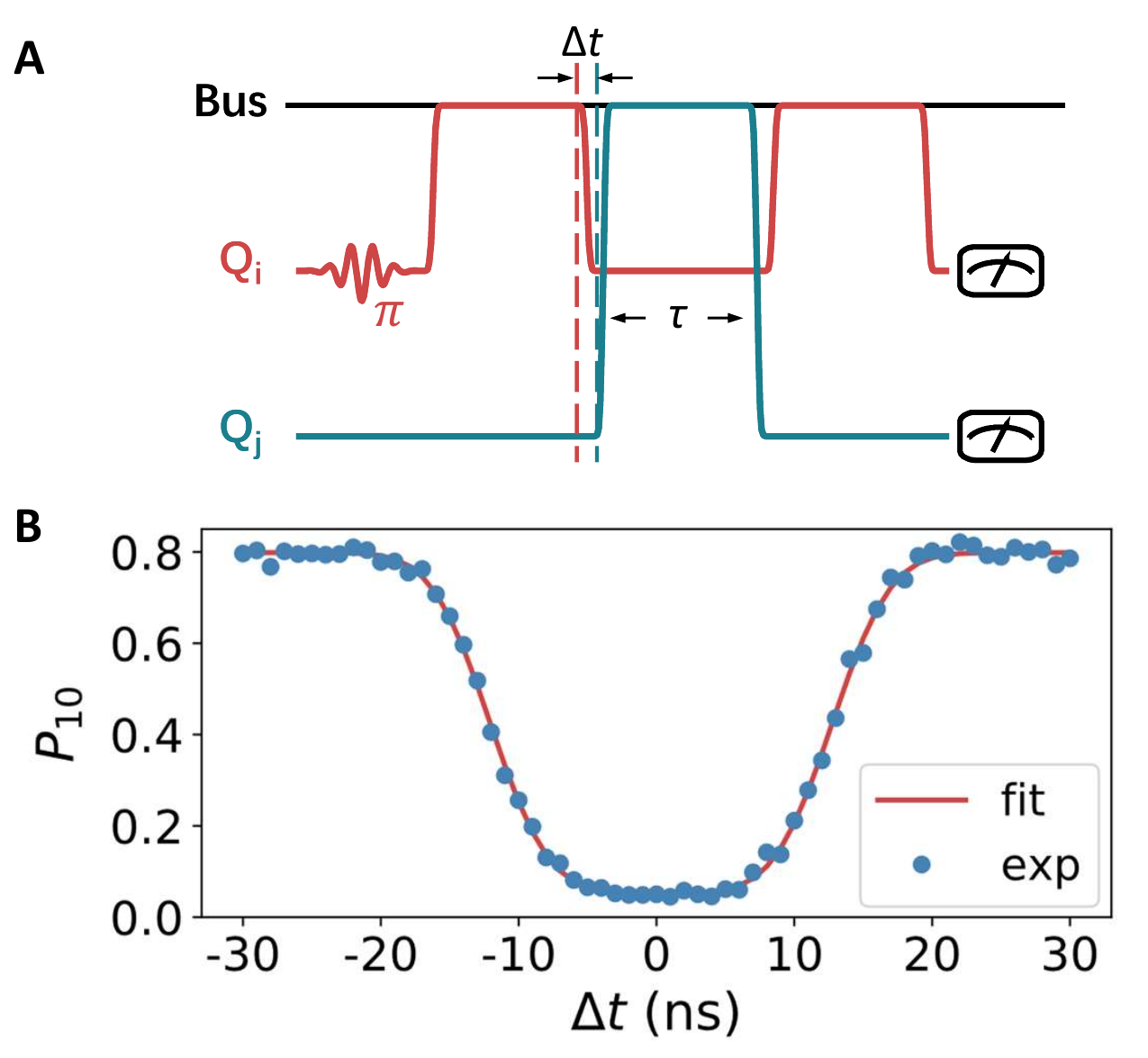}
\caption{{\bf Z pulse timing via the bus resonator.} ({\bf A}) Pulse sequence used to calibrate the Z pulse timing between $Q_i$ and $Q_j$. By varying the delay $\Delta t$, the two Z control lines for the qubit pair $\{Q_{i}, Q_{j}\}$ can be synchronized. When the relative delay $\Delta t$ is small, the microwave photon in $Q_{i}$ excited with its $\pi$ pulse will be swapped to $Q_{j}$ at the end of the sequence, yielding a small value of the measured probability in $Q_{i}$. In contrast, when $|\Delta t|$ is noticeable, $Q_{i}$'s photon will be swapped into the bus resonator and then back, with $Q_{i}$'s excitation remaining significant. ({\bf B}) Experimental data for a synchronized qubit pair $\{Q_{31}, Q_{1}\}$. Blue dots are experimentally measured probability $P_{10}$ using the pulse sequence in {\bf A}. The center of the well shape determines the timing offset $\Delta t$.}
\label{fig:timing}
\end{figure}

The pulse sequence for calibrating the Z pulse timing offset between the qubit pair $\{Q_i, Q_j\}$ is shown in Fig. \ref{fig:timing}(\textbf{A}). $Q_i$ is excited to $|1\rangle$ with a $\pi $ pulse and then two square pulses separated by a time delay $\tau$ for swapping the photon between the bus resonator $\cal {R}$ and $Q_{i}$ are applied. At the same time, a square pulse with a length of $\tau$ for swapping the photon between $\cal {R}$ and $Q_{j}$ is applied, whose start time varies with a delay of $\Delta t$ relative to the first square pulse on $Q_i$. If $Q_{j}$'s square pulse is right inbeween the two square pulses on $Q_{i}$, signals through the Z control lines of $Q_{i}$ and $Q_{j}$ are well synchronized. The representative experimental data for a well synchronized qubit pair are shown in Fig. \ref{fig:timing}(\textbf{B}), which calibrates the Z pulse timing between the far separated $Q_{31}$ and $Q_{1}$.

\section{Dynamics of two-body correlations}
\begin{figure*}[h]
\includegraphics[width=1.8\columnwidth]{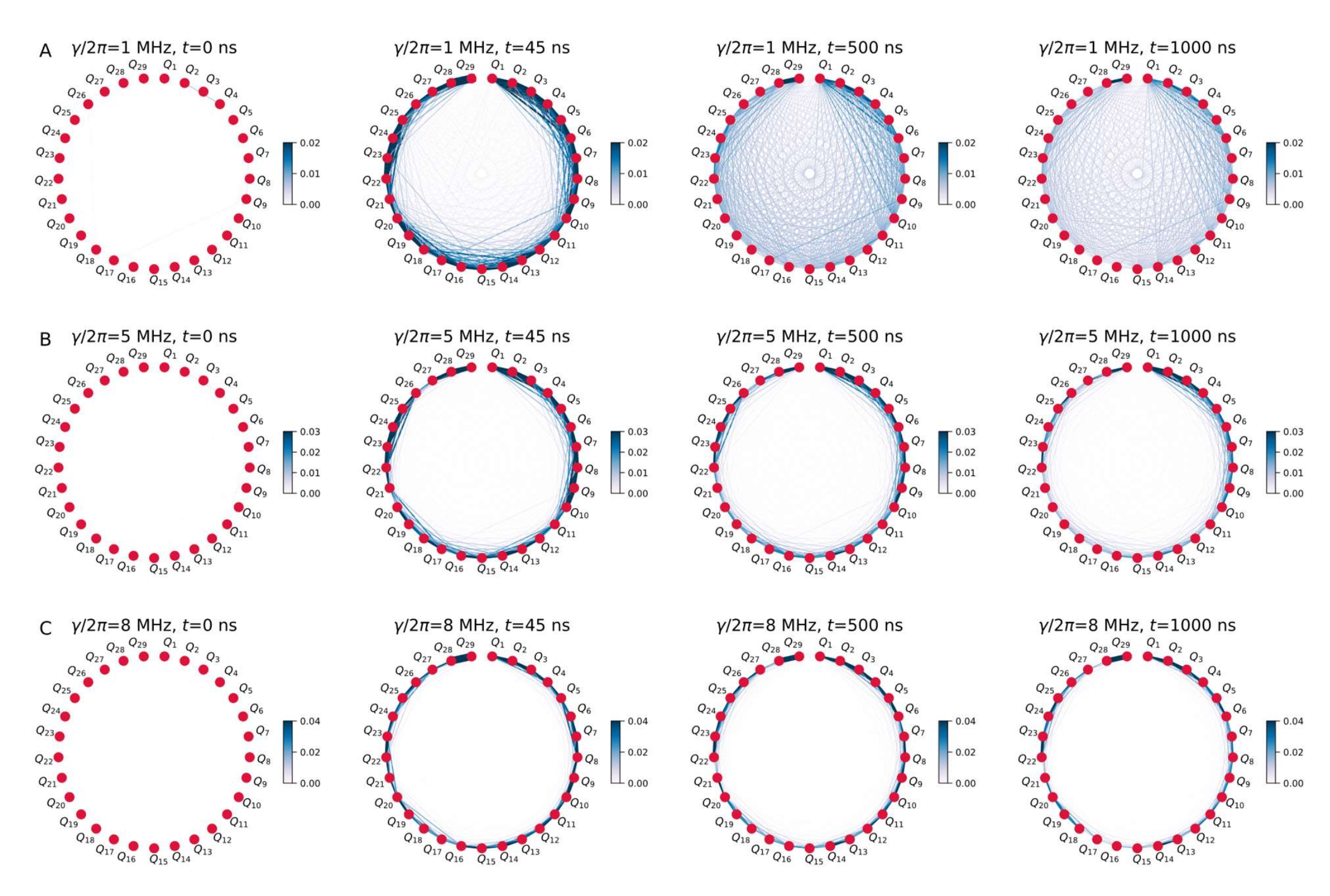}
\caption{{\bf Dynamics of two-body correlations.} Snapshots of the experimentally extracted two-body correlations $C(i,j)$ for $\gamma/2\pi = 1$ (\textbf{A}), 5 (\textbf{B}), and 8 MHz (\textbf{C}), from the initial time $t=0$, where the product state is initialized, to the largest experimental time, $t=1000$\ ns. A small Stark potential is not sufficient to prevent the build-up of correlations across the device, resulting in a largely entangled evolved state at long times. On the other hand, $\gamma/2\pi=5$ or 8 MHz, show the onset of non-ergodicity, where correlations are short- ranged with marginal growth within the experimentally accessible times.}
\label{fig:cij_dynamics}
\end{figure*}

One of the main characteristics of quantum many-body systems in out-of-equilibrium is the presence of large entanglement among its constituents. As we have argued in the main text, this can be inferred by the large degree of correlations between physically distant qubits, and owing the null-entangled initial product state, global entanglement builds up over the course of the dynamics. To better understand this evolution, we show in Fig.~\ref{fig:cij_dynamics} snapshots of the experimentally measured two-body correlations, further contrasting it with growing Stark potentials. Under the ergodic regime, $\gamma/2\pi = 1$\ MHz, the correlations quickly grow at very early times, in particular, for $t = 45$\ ns (around one tunneling time $\propto 1/\overline{J_{ij}}$), they extend already beyond nearest-neighbor qubits. Equilibration is already seen at $t=500$\ ns, with $C(i,j)$'s spread over all qubit pairs, in similarity to the results at the largest experimental time $t=1000$\ ns.

For larger tilt potentials ($\gamma/2\pi = 5$ and 8 MHz), the build-up of correlations at early times for nearest-neighbors similarly occurs, however, it quickly gets `frozen' in this regime, with no significant growth past this short-ranged scale within experimentally accessible times. This marks the non-ergodic behavior outlined in the main text, due to the presence of the nonrandom potential.

\section{Numerical details}

We focus on the dynamics at the middle of the eigenspectrum, thus one needs knowledge of the maximum and minimum energies in advance in order to select initial states (which univocally define the system's total energy) as to probe this region of the spectrum. In practice, the two extremal states are calculated by solving the ground state of $H$ and $-H$, via the density matrix renormalizaton group (DMRG) method. In this, we perform 10 DMRG sweeps to obtain a highly accurate target state, with truncation error smaller than $10^{-12}$.

To benchmark the real-time dynamics with experimental results, we adopted the time-dependent variational principle (TDVP) algorithm in finite matrix product states (MPSs)~\cite{Haegeman2011}. By projecting the time-dependent Schr\"odinger equation to the tangent space of the MPS manifold at each time step, this method is able to handle real-time dynamics of the system with long-ranged terms. However, for the evolution starting from a product state, which is a MPS of bond dimension one with a limited tangent space, both one-site and two-site TDVP methods may fail to capture the true direction of motion. In our calculations, we take the global subspace expansion procedure~\cite{Yang2020} in the first 20 time steps to enlarge the tangent space of the low entangled state. As to balance accuracy and computational complexity, we use more-expensive two-site TDVP algorithms in the first 100 time steps, and one-site sweeps in the subsequent ones, when the tangent space contains sufficient degree of freedom. For the same reason, various discrete time steps are selected from 2 to 5 nanoseconds, for different $\gamma$'s. The maximum bond dimension is 3000 for all calculations in the main text; and the truncation error is always smaller than $10^{-4}$.
\begin{figure}[h]
\includegraphics[width=1\columnwidth]{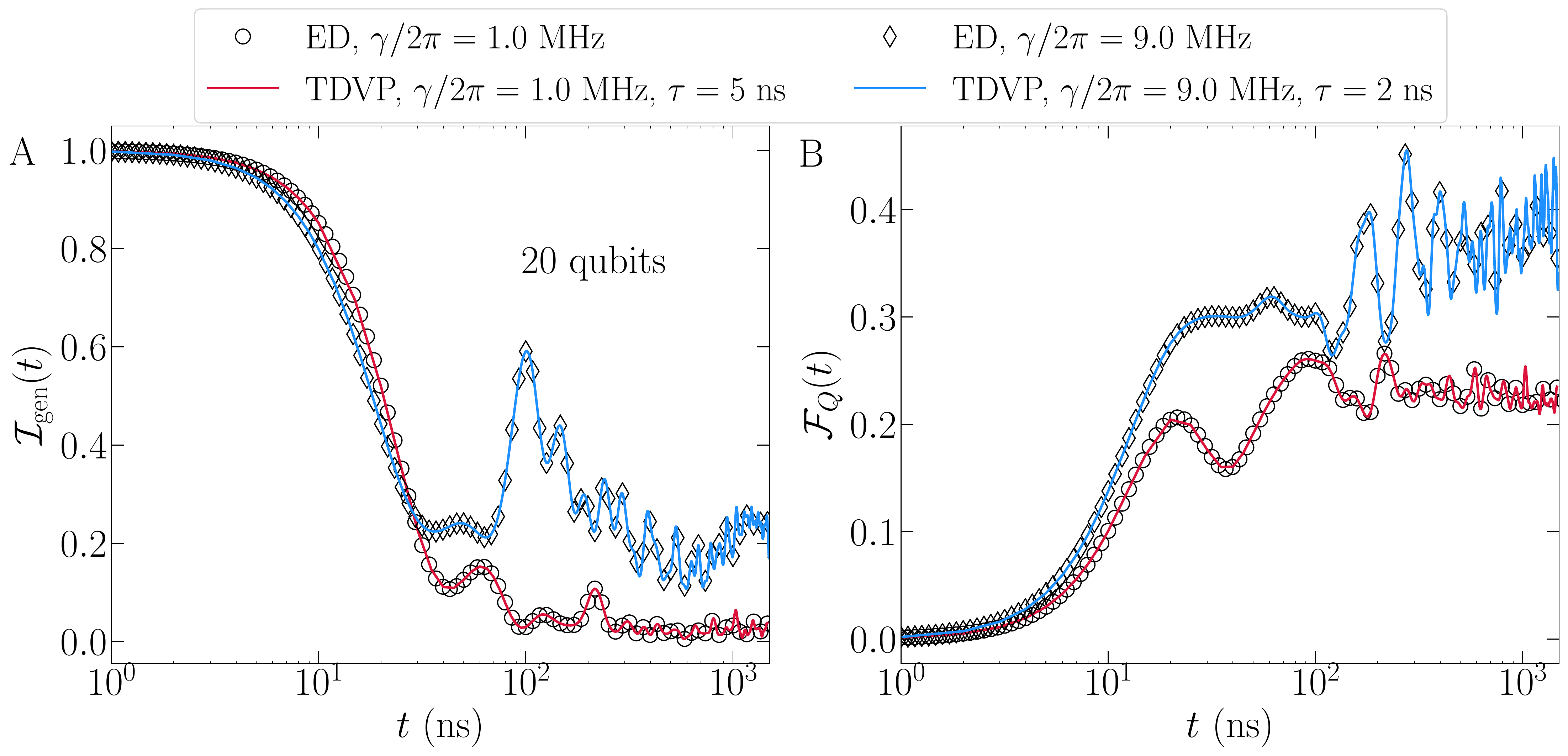}
\caption{{\bf TDVP benchmark against exact diagonalization (ED) results.} Numerically computed time-dependence of the generalized imbalance (\textbf{A}), and of the quantum Fisher information (\textbf{B}), when selecting 20 qubits of the device, preserving the experimentally relevant parameters. Markers (continuous lines) depict the ED (TDVP) results at small tilt potential $\gamma/2\pi = 1$\ MHz or large one $\gamma/2\pi = 8$\ MHz. In both cases, within the experimentally relevant times, the agreement between the different numerical methods is remarkable, and the evolution is studied from a single initial product state.}
\label{fig:benchmark_ed}
\end{figure}

\begin{figure}[h]
\includegraphics[width=1\columnwidth]{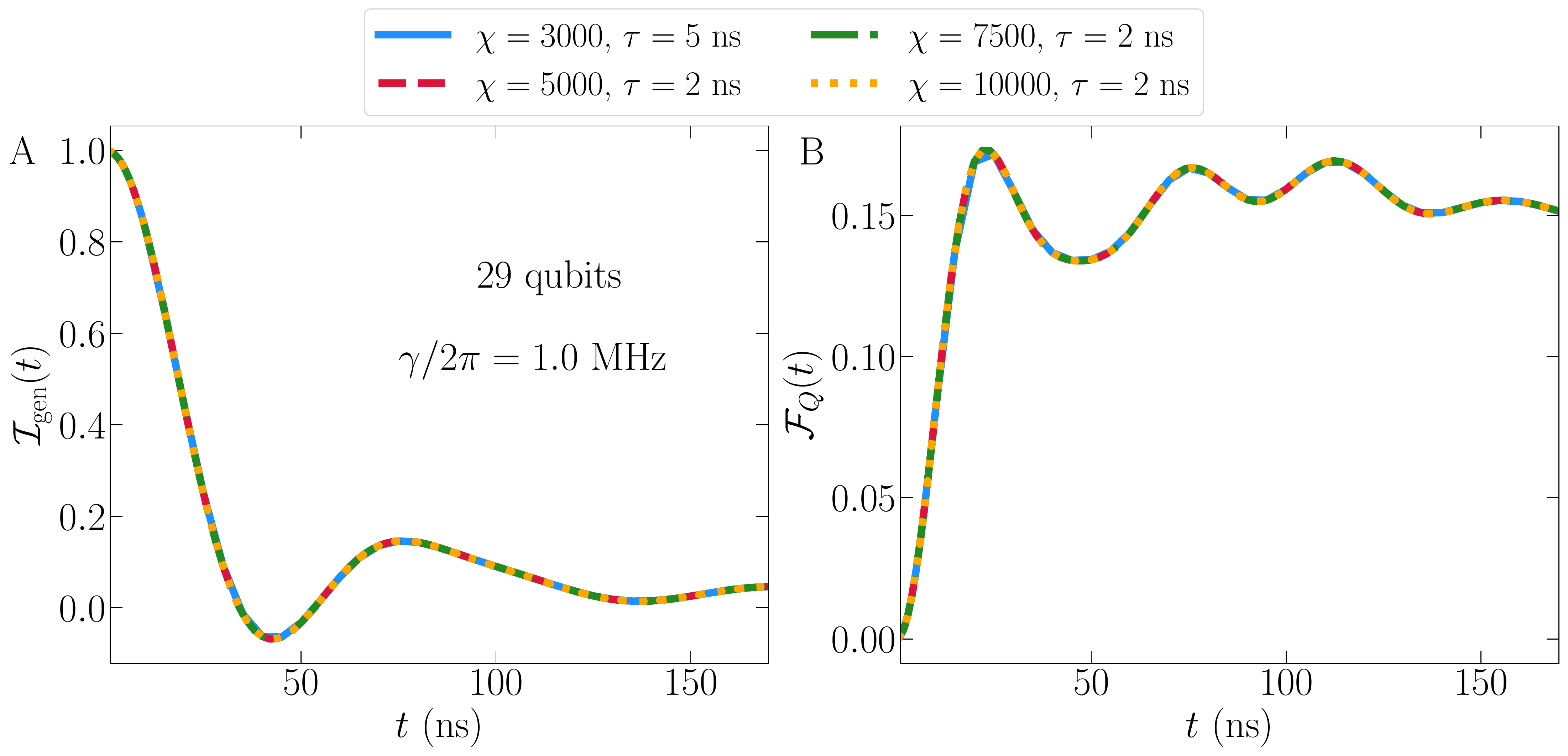}
\caption{{\bf Convergence of the TDVP evolution with bond-dimension $\chi$.} We systematically test the convergence of the dynamics within the TDVP numerical approach by computing the time-dependence of the generalized imbalance (\textbf{A}), and of the quantum Fisher information (\textbf{B}), when tracking all of the $N=29$ qubits experimentally used. Data is presented at the regime of largely ergodic behavior $\gamma/2\pi = 1$\ MHz (and much smaller than the average coupling $\overline{J_{ij}}$). Up until close to the fast equilibration times, dynamics for all bond-dimensions $\chi$ is yet indiscernible. As with the ED benchmark, we study the evolution based on a single initial product state.}
\label{fig:benchmark_tdvp}
\end{figure}

When using a smaller number of qubits ($N = 20$), with the experimentally relevant parameters, we have further benchmarked the evolution obtained via TDVP with the exact results of the unitary evolution (Fig.~\ref{fig:benchmark_ed}) within the experimental pertinent times. The agreement, either for small or large tilt potentials $\gamma$ is remarkable. Now, fully resorting to the total number of qubits experimentally employed ($N=29$), we further test the convergence of the TDVP results by systematically enlarging the bond-dimension $\chi$ (Fig.~\ref{fig:benchmark_tdvp}), at the most challenging regime of the parameter space, i.e., when $\gamma$ is small, and large entanglement is still manifest (See Fig.~\ref{fig:3} in the main text). Again the unitary dynamics does not present quantitatively relevant discrepancies, attesting the usage of $\chi=3000$ as a sufficient bond dimension in this demanding scenario.

\section{Initial state average}
For a system without explicit disorder, the standard procedure of averaging observed quantities over an ensemble of disorder realizations is no longer applicable. On top of usual measurement repetitions for a given initial state, we employ a second scheme that increases the overall statistics of the presented data. To start, we numerically obtain the spectrum of all product states, that is, $E^{(n)}=\langle\Psi_0^{(n)}|H|\Psi_0^{(n)}\rangle$ with a given conserved number of photon excitations, for a given $\gamma$ potential. Then, after using the previously explained scheme for numerically extracting the ground state energy $E_{\rm GS}$ of $H$ and its highest excited state $E_{\rm max}$, we are able to locate the product state energies in the eigenenergy density spectrum of the Hamiltonian.
\begin{figure}[h!]
\includegraphics[width=1\columnwidth]{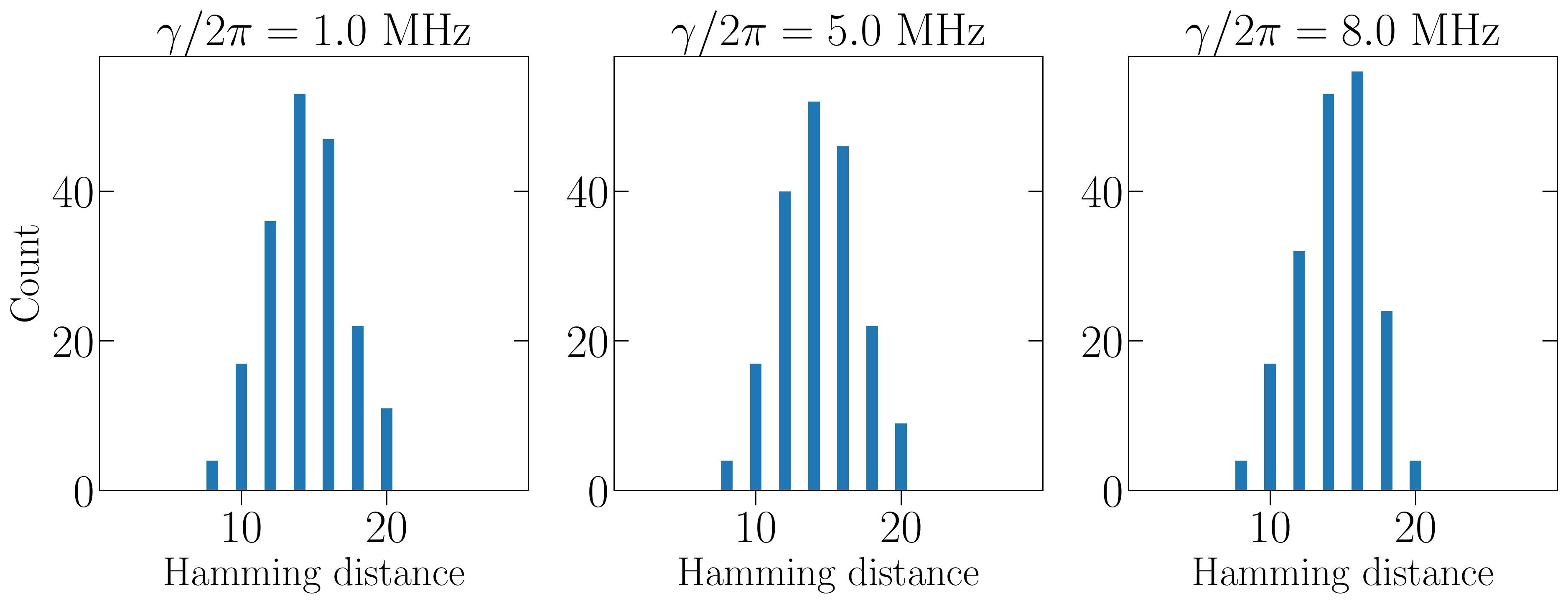}
\caption{{\bf Distribution of Hamming distances between pairs of initial states.} In order to promote statistical average with a deterministic potential, we average the extracted observables over 20 initial product states for each value of $\gamma$, whose energies lie in the center of the eigenspectrum of $H$. Their mutual Hamming distance is computed for $\gamma/2\pi = 1$ (left), 5 (center) and 8 MHz (right), and display an overall Gaussian profile; the same set of states is used in both experiments and numerical data.}
\label{fig:initial_states}
\end{figure}

We thus proceed by randomly selecting 20 initial product states within a narrow energy density window $\varepsilon_{\rm target}=0.5\pm0.02$. To understand how unrelated are these 20 initial states, we compute the standard Hamming distance $\cal HD$ (a measure of how different are two bitstrings representing the product states by counting the number of different bits, and not a dynamical observable as used in the main text) between each pair of states in this subset. Histograms of the $\cal HD$ for the initial states used are presented in Fig.~\ref{fig:initial_states}, over a large range of tilt potentials. The histogram profiles are essentially Gaussian within this approach, and allow one to use this procedure to promote statistical averages of the quantities in the main text.

\section{ETH analysis and the emerging dipole conservation}
To better understand the onset of non-ergodic behavior with growing tilt potential $\gamma$, we will now make use of typical ETH expectations, accompanied by quantum chaotic predictions on the level repulsion of eigenenergies $E_\alpha$ of $H$. For that, we will restrict the numerical analysis, to up to 18 qubits, in order to make it amenable to exact diagonalization methods, while keeping the experimentally relevant parameters. When promoting such system size analysis, the choice of qubits is made such as to denote the smallest `volume' in the corresponding lattice.

\begin{figure*}[t]
\includegraphics[width=1.9\columnwidth]{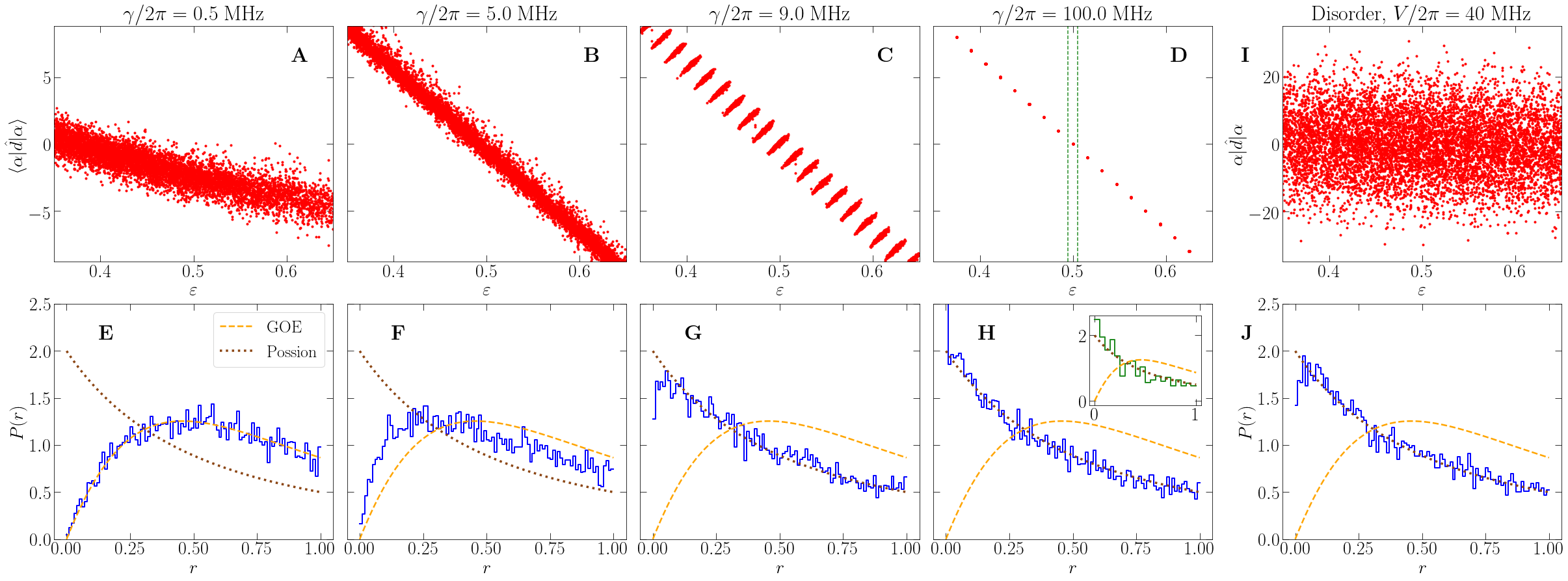}
\caption{\textbf{EEV's and ratio of adjacent gaps.} (\textbf{A})--(\textbf{D}) Numerically extracted eigenstate expectation values of the dipolar operator $\sum_j j \hat n_j$ close to the center of the spectrum ($\varepsilon = 0.5$) on a 16-qubit version of the device, with increasing potential $\gamma$ as indicated.  To facilitate visualization with the growing spectrum width as the Stark term is varied, we re-scale the eigenenergies $E_\alpha$ by plotting the energy density $\varepsilon = (E_\alpha - E_{\rm GS})/(E_{\rm max}-E_{\rm GS})$ instead, where $E_{\rm GS}$ is the Hamiltonian's ground state, and $E_{\rm max}$, its largest eigenenergy. (\textbf{E})--(\textbf{H}) Distribution of the ratio of adjacent gaps $r$ corresponding to the same tilt potentials as in (\textbf{A})--(\textbf{D}). As a direct comparison, (\textbf{I}) and (\textbf{J}) display the corresponding $\langle \alpha| \hat d|\alpha\rangle$ and $P(r)$ if instead the energy landscape $\{W_j\}$ is randomly selected from a uniform distribution $W_j\in [-V,V]$, as in a typical MBL process. Level repulsion is similarly lost, but dipole-moment conservation is no longer present. The dashed lines in (\textbf{D}) limit the energies of the fragment with zero dipole moment, and whose ratio of adjacent gaps is analyzed in the inset of (\textbf{H}): again, level repulsion that is not manifested over the whole spectrum, is still absent within each fragmented subspace.}
\label{fig:EEVs}
\end{figure*}

\begin{figure*}[t]
\includegraphics[width=1.5\columnwidth]{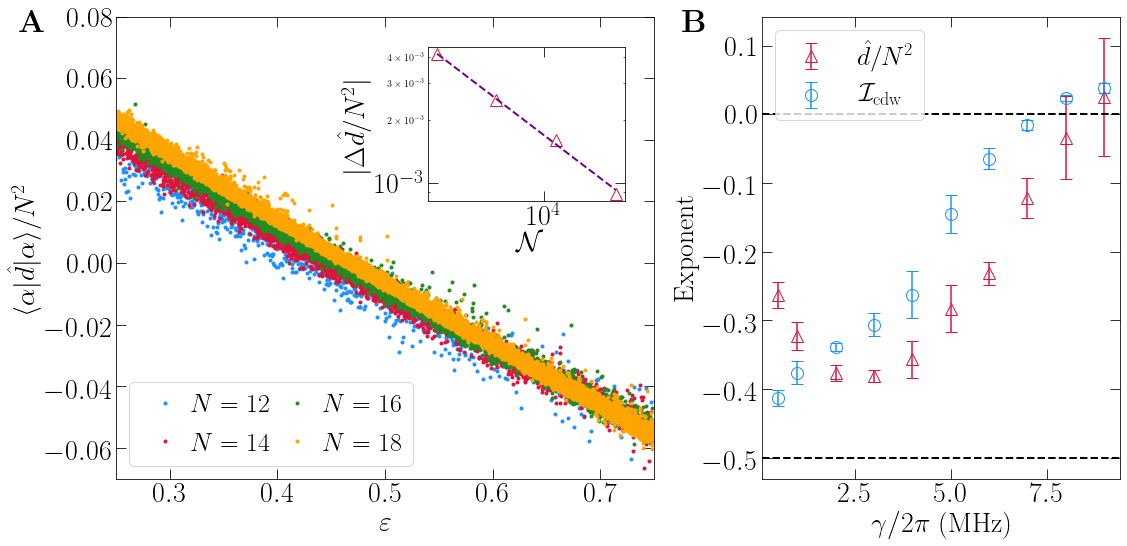}
\caption{\textbf{The fluctuations of consecutive EEVs.} (\textbf{A}) The EEVs of the normalized dipole-operator $(1/N^2)\hat d$ for different system sizes, ranging from $N=12$ to 18 qubits at $\gamma/2\pi = 2$\ MHz; inset displays the corresponding scaling of the fluctuations with the Hilbert size dimension for the data set in the main panel. (\textbf{B}) The dependence of the scaling exponent of the EEV fluctuations on the tilt potential slope, obtained for both the normalized dipole and the standard charge-density wave imbalance operator $\cal \hat I$. The ETH predicted value, and its complete breakdown, are marked by the horizontal dashed lines at -0.5 and 0, respectively. Error bars stem from fitting uncertainties.
}
\label{fig:EEVs_fluctuation}
\end{figure*}

With increasing Stark potential, the Hilbert space fragments in subspaces, where the dipole-moment emerges as a conserved quantity in the limit $\gamma\to\infty$. To test this, we show in Fig.~\ref{fig:EEVs} the eigenstate expectation values (EEV) of the dipole moment operator $\hat d\equiv\sum_j j \hat n_j$, when selecting 16 qubits. At small $\gamma$ values, the expectation values display characteristic features of a thermalizing system, featuring a smooth variation with the energy, and exponentially small fluctuations in the system size (See Fig.~\ref{fig:EEVs_fluctuation} for a quantitative analysis). Directly related to the quantum chaotic predictions, the distribution of the ratio of adjacent gaps $r_\alpha = \min(\delta_\alpha,\delta_{\alpha+1})/\max(\delta_\alpha,\delta_{\alpha+1})$, with consecutive gaps  $\delta_\alpha\equiv E_{\alpha+1} - E_{\alpha}$ in the ordered list of eigenenergies $\{E_\alpha\}$, follows the corresponding random matrix ensemble expectation (Gaussian orthogonal ensemble - GOE):  $P_{\rm GOE}(r) = \frac{27}{8}\frac{r+r^2}{(1 + r + r^2)^{\frac{5}{2}}}$~\cite{Atas13}, at $\gamma \lesssim \overline{J}$. Larger Stark potentials, on the other hand, remove the characteristic level repulsion and a Poisson distribution $P_{\rm P} (r) = 2/(1+r)^2$ ensues, preceding the appearance of primordial fragmentation. For $\gamma\gg \overline{J}$, full fragmentation develops, where each eigenstate subspace is separated by approximately $\gamma$, and is characterized by
an \textit{integer} dipole-moment quantum number. We further contrast these results with a standard MBL scenario [See Figs.~\ref{fig:EEVs}(\textbf{I}) and (\textbf{J})], where we randomly set the onsite energy landscape from a uniform distribution. In this case, level repulsion is similarly absent, but dipole-moment is no longer a conserved quantity.

A quantitative analysis of the validity of the ETH can be made by the scaling form of the average value of the eigenstate-to-eigenstate fluctuations with growing dimension of the Hilbert space~\cite{Alessio2016}. For that, we compute the absolute difference of EEVs in consecutive eigenstates, $\Delta {\cal O} \equiv |\langle\alpha|\hat {\cal O}|\alpha\rangle-\langle\alpha+1|\hat {\cal O}|\alpha+1\rangle|$, which provides an estimation of the fluctuations for a generic few-body operator $\hat {\cal O}$. Due to the extensive nature of the the dipole-moment operator, we analyze its normalized version, $(1/N^2)\hat d$, in Fig.~\ref{fig:EEVs_fluctuation}.

It has been shown for non-integrable models satisfying the ETH, that such fluctuations decay as a power-law of the Hilbert space dimension, ${\cal N}^{a}$, with $a=-1/2$ if sufficiently far from their integrable points~\cite{Beugeling14}. We test the scaling exponent for a range of values of the tilt potential in Fig.~\ref{fig:EEVs_fluctuation}(\textbf{B}), restricting it to the regime before fragmentation takes place [See Fig.~\ref{fig:EEVs}(\textbf{C})]. At small tilt-potentials, the scaling exponents are slightly above $-1/2$, which we attribute to the finite-size effects that can affect the scaling, even more due the absence of perfect homogeneity of the device. Nevertheless, the approach to zero with increasing $\gamma$ is a direct signature of the non-ergodic behavior, marking the breakdown of ETH. Moreover, if one instead  repeats this analysis by selecting the more standard generalized imbalance (see main text) as the few-body operator under study, we notice similar characteristics for the departure of the ETH predictions [Fig.~\ref{fig:EEVs_fluctuation}(\textbf{B}], ruling out the particularity of choosing an observable related to the emerging integrability at large $\gamma$'s.

Now, since the standard experimental characterization of Stark MBL is performed via the quench dynamics of initial product states, one needs a metric to quantify how many eigenstates effectively contribute to the evolution of few-body observables. This can be readily obtained by numerically computing the width in energy associated to the initial state in the eigenenergy basis $\sigma(E) \equiv [\langle H^2\rangle - \langle H\rangle^2]^{\frac{1}{2}}=[\sum_\alpha |\langle\Psi_0|\alpha\rangle|^2(E_\alpha - E)^2]^{\frac{1}{2}}$, for the energy conserving unitary evolution $E = \langle \Psi_0|H|\Psi_0\rangle = \langle \Psi_t|H|\Psi_t\rangle$. It is a time-independent quantity, reflecting the relevant Hilbert space spanned over the course of the dynamics.

\begin{figure*}[t]
 \includegraphics[width=2\columnwidth]{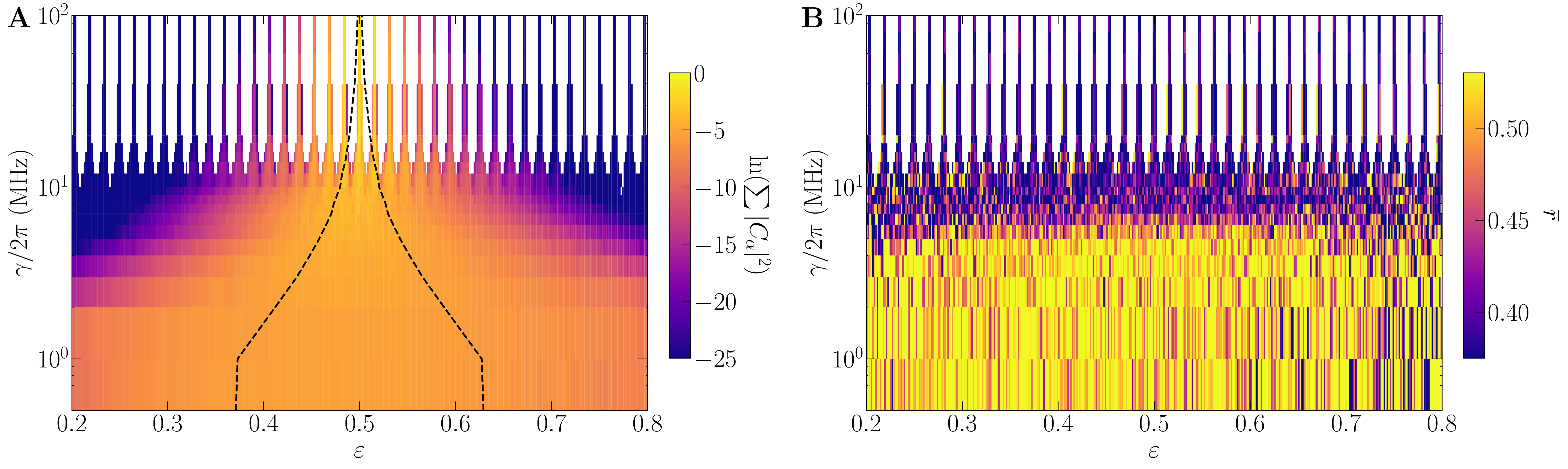}
 \caption{\textbf{The $\gamma$ vs $\varepsilon$ numerical phase diagram for 16-qubits.} (\textbf{A}) Color profile depicts the logarithm of the sum of probabilities $|C_\alpha|^2$ within each eigenenergy density window $d\varepsilon$. Dashed lines denote the average width $\sigma$ of the initial state $|\Psi_0\rangle$ in the Hamiltonian $H$, providing an estimation of the eigenstates effectively contributing to the dynamics.  As in Fig.~\ref{fig:EEVs}, we restrict the analysis to 16-qubits, and re-scale the energy axis by displaying the energy density $\varepsilon$ instead. (\textbf{B}) Average value of the ratio of adjacent gaps $\overline{r}$ in the same $\gamma -\varepsilon$ space: large tilt potentials yield in uncorrelated eigenenergies such that the $\overline{r}_{\varepsilon}$ within the small energy windows $d\varepsilon$ quickly drops to the Poisson value $r_{\rm P}\simeq0.39$, even before fragmentation takes place at larger $\gamma$'s.}
\label{fig:overlaps}
\end{figure*}

Figure~\ref{fig:overlaps}(\textbf{A}) presents the energy (density) resolved overlaps $|C_\alpha|^2\equiv|\langle\Psi_0|\alpha\rangle|^2$ for a large range of tilt potentials, when taking $|\Psi_0\rangle$ as an initial product state residing in the center of the eigenspectrum. As $\gamma$ increases, and fragmentation takes place at $\gamma/2\pi \simeq 10$\ MHz, a fairly small subset of eigenstate energies possess significant contribution on the overlaps, unlike in the ergodic regime at $\gamma \simeq \overline{J}$. This is the regime where one expects well developed Bloch oscillations as we experimentally demonstrate occurring at $\gamma/2\pi = 16$\ MHz in the main text. Nonetheless, we have also argued there that smaller Stark potentials are already sufficient to induce the breakdown of ergodicity, being signified by either local-memory persistence over the course of the dynamics or the presence of slow (logarithm-in-time) growth of entanglement. However, according to the quantum chaos predictions~\cite{Atas13,Alessio2016}, non-ergodic behavior is directly inferred by energy level correlation (repulsion or its absence), and can be quantified by the average value of the ratio of adjacent gaps, $\overline{r}_\varepsilon$, taken within small energy windows .

Figure~\ref{fig:overlaps}(\textbf{B}) displays a similar `phase diagram' of $\overline{r}_\varepsilon$ in the $\gamma-\varepsilon$ parameter space. At small $\gamma$'s, ergodic behavior is largely manifest, with ratio of adjacent gaps admitting values close to the GOE average $\overline r_{\rm GOE} = 4-2\sqrt{3} \approx 0.5359$. On the other hand, much before the onset of fragmentation, energy levels start to become uncorrelated at $\gamma/2\pi \simeq 5$\ MHz, and as a result, $\overline r_\varepsilon\to\overline{r}_{\rm P}\approx0.3862$. This goes in line with estimations of the onset of Stark MBL in the main text, although here using a much smaller Hilbert space, in order to make the extraction of exact eigenpairs amenable.

\end{document}